\documentclass[12pt]{article}
\usepackage[utf8]{inputenc}
\usepackage[russian,english]{babel} 
\usepackage[T2A]{fontenc}   
\usepackage{mathtools}
\usepackage{amsfonts,amssymb,amsmath} 
\usepackage{graphicx} 
\usepackage{cite}
\usepackage{hyperref}
\hypersetup{colorlinks=true, linkcolor=blue, filecolor=blue, urlcolor=blue} 

\newtheorem{statement}{Statement}

\allowdisplaybreaks

\begin{document}

\title{Optimization of Time-Dependent Decoherence Rates and Coherent Control for a~Qutrit System}

\author{ Oleg~V.~Morzhin$^{1,2,}$\footnote{E-mail: \url{morzhin.oleg@yandex.ru};~ 
    \href{http://www.mathnet.ru/eng/person30382}{mathnet.ru/eng/person30382};~ 
    \href{https://orcid.org/0000-0002-9890-1303}{ORCID 0000-0002-9890-1303}} 
    \quad and \quad 
Alexander~N.~Pechen$^{1,2,}$\footnote{E-mail: \url{apechen@gmail.com};~ 
    \href{http://www.mathnet.ru/eng/person17991}{mathnet.ru/eng/person17991};~ 
    \href{https://orcid.org/0000-0001-8290-8300}{ORCID 0000-0001-8290-8300}} 
    \vspace{0.2cm} \\
$^1$ Department of Mathematical Methods for Quantum Technologies,\\
    Steklov Mathematical Institute of Russian Academy of Sciences, \\
8 Gubkina St., Moscow, 119991, Russia;\\
$^2$ Quantum Engineering Research and Education Center, \\
    University of Science and Technology MISIS,\\
6 Leninskiy Prospekt, Moscow, 119991, Russia}

\date{}
\maketitle

\begin{abstract}
The work considers an open qutrit system whose density matrix $\rho(t)$ evolution is  governed by the Gorini--Kossakowski--Sudarshan--Lindblad master equation
with simultaneous coherent (in the Hamiltonian) and incoherent
(in the superoperator of dissipation) controls. Incoherent control makes the decoherence rates depending on time in a specific controlled manner and within clear physical mechanics. We consider the problem 
of maximizing the Hilbert--Schmidt overlap between the system's final state $\rho(T)$ and a~given target state $\rho_{\rm target}$ and the problem of minimizing the squared Hilbert--Schmidt distance between these states. For the both problems, we perform their realifications, derive the corresponding Pontryagin function, adjount system (with the two cases of transversality conditions in view of the two terminal objectives), and gradients of the objectives, adapt the one-, two-, three-step gradient projection methods. 
For the problem of maximizing the overlap, we also adapt the regularized first-order Krotov method. In the numerical experiments, we analyze, first, the methods' operation and, second, the obtained control processes, in respect to 
considering the environment as a resource via incoherent control.
\vspace{0.3cm}

\end{abstract} 

\rightline{\it Dedicated to the 80th anniversary}
\rightline{\it of the academician Alexander S. Holevo} 

\section{Introduction} 

The theory of quantum information~\cite{Holevo_Book_2010_2019, Holevo_Book_2020_2011, Wilde_Book_2017}
is the quantum extension of the conventional theory of information, where the concept of bit is extended to the concept of 
qubit, the Shannon entropy is replaced by the von Neumann entropy, etc.
The theory of (optimal) control of quantum systems 
such as atoms, molecules, etc., (e.g.,~\cite{KochEPJQuantumTechnol2022, ButkovskiyBook_1984_1990,
TannorBook2007, BrifNewJPhys2010, DongPetersen2010,
MooreCS2011, Gough2012, DongWuYuanLiTarn2015,
KochJPhysCondensMatter2016, DAlessandroBook2021,
BaiChenWuAn2021, CastroGiovanniniEtAlNJP2023, KuprovBook2023}) is the quantum extension of 
the conventional theory of (optimal) control~\cite{PontryaginBook1962, KrotovBook1996}.
As it is known, modeling of various quantum information's entities (generation of quantum gates, 
transfers along a spin chain) is carried out in terms of optimal control problems for various
quantum mechanical equations with controls under various corresponding fidelities, constraints
(e.g., \cite{PalaoKosloffPRL2002, BochkinFeldmanLazarevPechenZenchuk2022}). 

Control of various quantum systems is an important area at the intersection of mathematics, 
physics, chemistry, and computer science, which is crucial for development 
of modern quantum technologies. Considering a~quantum system whose dynamics is governed by some quantum mechanical equation 
(e.g., Schr\"{o}dinger, von Neumann, Gorini--Kossakowski--Sudarshan--Lindblad (GKSL), etc.) depending 
on control which represents certain external influence, e.g., laser light and/or certain 
characteristics of the system's environment, one is interested to analyze the evolution equation from the point of view of
what possibilities of control there are in regard to certain objectives and constraints, in addition to the possibilities 
that the quantum system has without controls.

The theory of optimal control for quantum systems is actively developed since 1980s. 
For example, the article \cite{KazakovKrotov1987} (1987) considers some open three-level 
quantum system with coherent control, where the goal is to maximize the system's inverse population; the article~\cite{PeirceDahlehRabitzPRA1988} (1988) is devoted to optimal coherent control for a~diatomic molecule described via an infinite-dimensional system. Nowadays,
a~large number of directions, papers, and books are devoted 
to quantum control. Various types of quantum systems and controls 
are considered, for example, infinite-dimensional open quantum systems with
coherent quantum feedback~\cite{GoughRatiuSmolyanov2021}, Floquet engineering~\cite{BaiChenWuAn2021, CastroGiovanniniEtAlNJP2023} for control of quantum systems using time-periodic external fields.  The~first experimental realization of Kraus maps studied in~\cite{Wu_2007_5681} for an open single qubit was done in~\cite{Zhang_Saripalli_Leamer_Glasser_Bondar_2022}. Developing quantum optimal control methods is actual for a further research. 

An important direction is quantum control of open quantum systems. An open quantum system is a quantum system immersed in some environment. On one 
hand, the system's environment is considered as {\it obstacle}. For example, the works 
\cite{Goerz_NJP_2014, FonsecaFanchiniLimaCastelano2022} 
are devoted to the problem of generating unitary quantum gates when quantum dynamics is dissipative.
On another hand, as the articles~\cite{PechenPRA062102.2006, PechenPRA2011} show, one can constructively and efficiently use the system's environment as a~control {\it resource}. In this regard, incoherent control by the environment is introduced into the corresponding quantum mechanical equation: first, in the superoperator of dissipation and, second, in the Hamiltonian taking into account the Lamb shift --- along with possible presence of coherent control in the Hamiltonian. In regard to the latter, the article~\cite{DannTobalinaKosloffPRA2020} considers coherent control entering both in the Hamiltonian and the superoperator of dissipation. Based on 
the articles~\cite{PechenPRA062102.2006, PechenPRA2011} (2006, 2011), 
the further works~\cite{MorzhinPechenLJM2019, MorzhinPhysPartNucl2020, LokutsievskiyJPA2021, MorzhinPechenIJTP2021, 
PetruhanovPhotonics2023, MorzhinPechenQIP2023}, etc. consider various quantum systems with 
simultaneous coherent (in the Hamiltonian) and incoherent (in the effective Hamiltonian via the Lamb shift and in the superoperator of dissipation), various objective functionals, classes of controls (piecewise continuous, piecewise constant, etc.), constraints, and optimization methods. From physical point of view, such incoherent control should be non-negative. The work~\cite{KozyrevPechenPRA2022} is devoted to quantum feedback coherent control in quantum photosynthesis under decoherence. 

Various {\it three-level} quantum systems ($\Lambda$-atom, $V$, etc.) depending on control are studied, for example, 
in the articles~\cite{KazakovKrotov1987, LikhanskiiNapartovich1982, LiPengPRA1985, WangMullerEtAl2005,
AccardiKozyrevPechen2006, SugnyKontz2008, JieYaoMinHaiRui2009,
ZobovShauro2011, PechenTannorPRL2011, PechenTannor2012PRL,
PechenTannor2012IsraelJChem, Mortensen2018, DAlessandroShellerZhu2020,
PetiziolArimondoEtAl2020, DehaghaniLoboPereiraAguiar2022,
XuSongWangYe2022, KorashyAxioms2023, LenziGabrick_et_al_QuantumReports2023,
ElovenkovaPechenQuantumReports2023,KuznetsovPechenLJM2023}. 
The term ``qutrit'' means a carrier of quantum information which is 
realized by a~three-level quantum system. This is an analogy with
the terms ``qubit'' and ``qudit'' meaning such carriers of quantum 
information which are realized, correspondingly, by two-level 
and $d$-level quantum systems. Coherent quantum control of a three-level $\Lambda$-atom 
interacting with the environment in the stochastic limit was investigated 
in~\cite{AccardiKozyrevPechen2006}. In~\cite{PechenTannorPRL2011, PechenTannor2012PRL}, control of a~three-level $\Lambda$-atom was studied for the analysis of quantum control landscapes, 
were, for this quantum system, a~second order trap was found. However, for $\Lambda$-atom second-order trap was shown to be not 
a~full trap~\cite{PechenTannor2012IsraelJChem}. The article~\cite{JieYaoMinHaiRui2009} 
is devoted to realizating two-qutrit quantum gates using certain control pulses for unitary dynamics. 

For closed and open quantum systems, various optimization tools were adapted and used including
the following (this list is not complete):
\begin{itemize}
    \vspace{-0.3cm}
    \item the Pontryagin maximum principle (see, e.g., \cite[Ch.~4]{ButkovskiyBook_1984_1990} 
and the survey~\cite{BoscainPRXQuantum2021}; in the theory of optimal control, this condition for optimality
is known for a~long time, see, e.g., the fundamental book~\cite{PontryaginBook1962}); 
    \vspace{-0.3cm}
    \item Krotov method~\cite{KrotovBook1996, TannorBook2007, KochJPhysCondensMatter2016,
KazakovKrotov1987, Goerz_NJP_2014,  
FonsecaFanchiniLimaCastelano2022, Tannor1992, Schirmer_deFouquieres_2011, MorzhinPechenUMN2019,
ArakiNoriGneitingPRA2023, MorzhinPechenIrkutsk} (beyond 
quantum control the foundation for this method was done in~\cite{KrotovFeldmanIzvAN1983}; 
see also Ch.~1 in the book~\cite{SrochkoVA_Book2000} with respect to the projection methods of the general Krotov method's type);
    \vspace{-0.3cm}
    \item gradient methods operating in a~functional space of controls: e.g., in the article~\cite{PeirceDahlehRabitzPRA1988} (1988); 
our works~\cite{MorzhinPechenLJM2019, MorzhinPechenIJTP2021, MorzhinPechenQIP2023, MorzhinPechenIrkutsk} 
use the gradient projection method's versions (GPM) and conditional gradient method for some open quantum systems with coherent and incoherent controls; the work~\cite{LyakhovPechen2022} uses gradient ascent search for some problem with the Schr\"{o}dinger equation and unconstrained coherent control; 
    \vspace{-0.3cm}
    \item gradient approach operating in a~finite-dimensional search space 
of some piecewise constant controls: e.g., GRadient Ascent Pulse Engineering (GRAPE)~\cite{KhanejaJMagnReson2005, SchulteHerbruggenSporlKhanejaGlaser2011, 
PechenTannor2012IsraelJChem, VolkovJPA2021, PetruhanovPechenIntJModPhysB2022, PetruhanovPhotonics2023, 
GoodwinVinding2023}, in particular, together with the Broyden--Fletcher--Goldfarb--Shanno (BFGS) method; 
our work~\cite{MorzhinIzvRAN2023} considers GRAPE and also finite-dimensional GPM;
    \vspace{-0.3cm}
    \item Chopped RAndom Basis (CRAB)~\cite{CanevaPRA2011, 
MullerSaidJelezkoCalarcoMontangero2022};
    \vspace{-0.3cm}
    \item speed gradient method~\cite{PechenBorisenokFradkov2022} (also see the survey~\cite{AndrievskyFradkovAiT2021}); 
    \vspace{-0.3cm}
    \item stochastic free-gradient tools oriented to approximate global finite-dimensional optimization,
for example, genetic algorithm~\cite{PechenPRA062102.2006}, 
dual annealing~\cite{VolkovJPA2021, MorzhinPechenIrkutsk}, etc. 
    \vspace{-0.1cm} 
\end{itemize}

In optimization, the notion ``projection'' is met in different senses. 
For optimizing control in the Shr\"{o}odinger equation, the works
\cite{ShaoCombesHauserNicotraPRA2022, ShaoNarisHauserNicotra2023} 
describe a~Newton method that uses a~projection operator to 
incorporate the dynamic equation directly into the objective, 
in particular, for a~$\Lambda$ system.

In this article, we consider an open qutrit system whose dynamics is governed 
by the GKSL master equation
with simultaneous coherent (in the Hamiltonian) and incoherent
(in the superoperator of dissipation) piecewise continuous and bounded controls. 
For this system, we consider the two objectives of Mayer type defined 
in terms of the Hilbert--Schmidt overlap and distance between 
the final $\rho(T)$ and predefined $\rho_{\rm target}$. 
Here we adapt the one-, two-, and three-step gradient 
projection methods (GPM-1, GPM-2, and GPM-3) using the 
works~\cite{LevitinPolyak1966, DemyanovRubinovBook1970,
PolakBook1971, PolyakUSSRComputMathMathPhys1964,
PolyakBook1987, AntipinDifferEqu1994,
VasilievNedic1994, NedichIzvVUZov1993,
VasilievNedic1993} and, when 
the objectives are linear in~state, 
the regularized in controls first-order Krotov method (RKM) is also used.

As in~\cite{FonsecaFanchiniLimaCastelano2022}, 
this work also considers some  open qutrit system using some form of the first-order RKM, 
but, in contrast to~\cite{FonsecaFanchiniLimaCastelano2022}, the considered by us 
qutrit open system contains both coherent and incoherent controls, we take arbitrary 
initial and target density matrices, the linear objective is the 
Hilbert--Schmidt overlap, and our RKM version takes into account 
the possibility to have, first, lower and upper magnitude constraints for coherent 
control and, second, upper magnitude constraint for incoherent control, 
in addition to the requirement for incoherent control to be non-negative. 

An important direction in the theory of optimization is acceleration of gradient methods.
A~fundamental contribution in this direction is the work~\cite{PolyakUSSRComputMathMathPhys1964}
(B.T.~Polyak, 1964) which develops the heavy-ball method as a~two-step (i.e. two previous
iterations' results are used for new iteration) gradient method
for unconstrained optimization. GPM-2 is related to the heavy-ball method and is intended 
for constrained optimization. GPM-3 continues this direction.

Another important direction in the theory of optimization is creation of various 
self-adaptive optimization algorithms; in this regard, see, e.g., 
the  article~\cite{Kaewyong_Sitthithakerngkiet_Axioms_2021}. 
A special interesting aspect of our article is that the optimization methods are used with the tuning parameters  fixed for the whole number of iterations.

Various representations are used for describing quantum states, e.g.,~\cite{KozlovSmolyanov2021, GoughOrlovSakbaevSmolyanov2021}. 
In contrast to~\cite{Goerz_NJP_2014, FonsecaFanchiniLimaCastelano2022, 
MorzhinPechenIrkutsk}, this article initially derives certain realification 
of the quantum system and objective functionals via some parameterization
of density matrix~$\rho$, and after that derives the methods' constructions 
for these realifications.  

The rest of our paper is organized as follows. Sec.~\ref{Section2} shortly describes the considered concept of incoherent control. Sec.~\ref{Section3} formulates the qutrit open system, objective functionals, constraints, and the realifications of these control problems. For these realificated problems, Sec.~\ref{Section4} 
adapts the various-step GPMs and the first-order RKM. Sec.~\ref{Section5} describes the corresponding numerical results. Conclusions Sec.~\ref{Section_Conclusion} resumes the article. 

\section{Incoherent Control with Time-Dependent Decoherence Rates}
\label{Section2}

The idea to use time-dependent decoherence rates adjusted by incoherent control was proposed in 2006 in~\cite{PechenPRA062102.2006}. In that work, master equations of the form 
\begin{equation*}
\frac{d\rho_t}{dt}=-i[H_c(t),\rho_t]+\sum\limits_k \gamma_k(t) {\cal D}_k,
\end{equation*}
where $H_c(t)$ is some controlled Hamiltonian, $\gamma_k(t)$ are controlled generally time-dependent decoherence rates, and ${\cal D}_k$  are some dissipators, were introduced and studied for quantum control with the key property of considering not some abstract dissipators  ${\cal D}_k$ but two particular physical forms of the GKSL dissipators derived in 1970th and 1980th in the weak coupling and low density limits. These forms correspond to two cases when Markovian dynamics of an open quantum system can be derived exactly and rigorously from exact microscopic interaction between quantum system and its environment, and physically correspond to interaction of the quantum system with photons (weak coupling limit) or with quantum gas via collision-type interaction (low density limit). In~\cite{PechenPRA062102.2006}, these Markovian master equations were modified to include the time-dependent decoherence rates $\gamma_k(t)$ tailored within physical models  via time-dependent spectral density of the environment $n(\omega,t)$ which described distribution of particles of the environment in their energy or frequency $\omega$ (or in other parameters characterizing states of environmental particles). This control was called as {\it incoherent control}. By the physical meaning as density, incoherent control satisfies $n(\omega,t)\ge 0$. Explicit expressions for $\gamma_k(t)$ in terms of $n(\omega,t)$ were provided in~\cite{PechenPRA062102.2006}. In addition to general consideration, two physical classes of the environment were exploited --- incoherent photons and quantum gas. In~\cite{PechenPRA2011}, it was shown that for the master equation with dissipators ${\cal D}_k$ derived in the weak coupling limit (describing for example an atom interacting with photons) generic $N$-level quantum systems become approximately completely controllable in the set of density matrices. The low density limit case with collisional-type decoherence is less studied in this context, although it corresponds to  the quantum linear Boltzmann equation~\cite{VacchiniPR2009} and describes for example a test particle in a quantum gas~\cite{VacchiniPRE2001}. A more detailed description of the concept of incoherent control is also provided in Sec.~2 of~\cite{MorzhinPechenQIP2023}. 

\section{Quantum System, Objective Functionals, Constraints, and Realification}
\label{Section3}

For a qutrit, the Hilbert space is $\mathcal{H} =  \mathbb{C}^3$. 
Consider the following GKSL master equation with coherent and incoherent controls:
\begin{align}
\dot\rho(t) = -i \big[H_0 + H_{u(t)}, \rho(t) \big] + \mathcal{L}^D_{n(t)}(\rho(t)), 
\quad \rho(0) = \rho_0, \quad t \in [0,T], 
\label{GKSL_system} 
\end{align}
where $\rho(t) : \mathcal{H} \to \mathcal{H}$ is  density matrix, i.e. positive semi-definite ($\rho(t) \geq 0$) with unit trace (${\rm Tr}\rho(t) = 1$), $\rho_0$ is a given initial density matrix; 
$u$ is scalar coherent control and $n=(n_1, n_2)$ is incoherent control; Hamiltonians $H_0$ and $H_{u(t)}$ determine, correspondindgly, 
free dynamics and interaction with $u(t)$; $\mathcal{L}^D_{n(t)}(\rho(t))$ is the controlled 
superoperator of dissipation which acts to $\rho(t)$. The full control $c=(u,n)$ is considered 
as piecewise continuous on $[0, T]$.  The final time~$T$ and initial density matrix $\rho_0$ are given. The system of units is such that the Planck's constant $\hbar=1$. As in \cite{PechenTannorPRL2011}, consider $\Lambda$-atom which is characterized by
$H_0 = \begin{pmatrix}
	0 & 0 & 0 \\
	0 & E_2 & 0 \\
        0 & 0 & E_3
\end{pmatrix}$ and $V = \begin{pmatrix}
	0 & 0 & V_{13} \\
	0 & 0 & V_{23} \\
        V_{13}^{\ast} & V_{23}^{\ast} & 0
\end{pmatrix}$, where $V_{13} = V_{13}^{\ast}$, $V_{23} = V_{23}^{\ast}$ are assumed. 

The transitions 
$1 \leftrightarrow 3$ and $2 \leftrightarrow 3$ are admissible, but the transition 
$1 \leftrightarrow 2$ is forbidden. Since there are only two allowed transitions with transition frequencies $\omega_{13}=E_3$ and $\omega_{23}=E_3-E_2$, in the resonant approximation only two components of incoherent control $n(\omega,t)$, namely $n_1(t)=n(\omega_{13},t)$ and $n_2(t)=n(\omega_{23},t)$, contribute to the dynamics. 

The superoperator of dissipation acts on $\rho(t)$ as
\begin{align*}
\mathcal{L}^D_{n(t)}(\rho(t)) :=&  \sum\nolimits_{j=1}^2 C_{j,3} \,(n_j(t)+1) 
\left( 2 A_{j,3} \, \rho(t) \,A_{j,3}^{\dagger} - \{ A_{j,3}^{\dagger} \, A_{j,3}, \rho(t) \} \right) + \nonumber \\\
&+ C_{j,3} \, n_j(t) \left( 2 A_{j,3}^{\dagger} \, \rho(t) A_{j,3} - \{ A_{j,3} \, A_{j,3}^{\dagger}, \rho(t) \} \right), 
\end{align*}
and the admissible transitions are determined by the matrices 
$A_{j,3} = V_{j,3} \, | j \rangle \langle 3|$, $j=1,2$, i.e.
$A_{13} = V_{13} \, | 1 \rangle \langle 3| = 
\begin{pmatrix}
0 & 0 & V_{13} \\ 
0 & 0 & 0 \\
0 & 0 & 0 
\end{pmatrix}$, $A_{13}^{\dagger} = 
\begin{pmatrix}
0 & 0 & 0 \\ 
0 & 0 & 0 \\
V_{13} & 0 & 0 
\end{pmatrix}$, 
$A_{23} = V_{23} \, | 2 \rangle \langle 3| = 
\begin{pmatrix}
0 & 0 & 0 \\ 
0 & 0 & V_{23} \\
0 & 0 & 0 
\end{pmatrix}$, 
$A_{23}^{\dagger} = 
\begin{pmatrix}
0 & 0 & 0 \\ 
0 & 0 & 0 \\
0 & V_{23} & 0 
\end{pmatrix}$. 
The physical meaning of controls $n_1,~n_2$ as densities of incoherent photons requires the constraints
\begin{align}
n_j(t) \geq 0, \quad j = 1,2, \quad t \in [0, T].
\label{required_constraint_for_n_j}
\end{align}
In addition, one can consider the constraints
\begin{align}
-\mu \leq u(t) \leq \mu, \quad \mu > 0, \quad n_j(t) \leq n_{\max}, \quad j = 1,2, \quad n_{\max}>0, \quad t \in [0, T].
\label{additional_constraints_for_controls}
\end{align}
If only (\ref{required_constraint_for_n_j}) is used, then we consider
$Q_{\infty} := \mathbb{R} \times [0, \infty)^2$ and $c(t) \in Q_{\infty}$.
If~both (\ref{required_constraint_for_n_j}) and (\ref{additional_constraints_for_controls}) are used, then we consider $Q_{\rm comp} := [-\mu, \mu] \times [0, n_{\max}]^2$ and $c(t) \in Q_{\rm comp}$.

For (\ref{GKSL_system}), consider the following objective functionals of the Mayer form:
\begin{align}
J_1(c) &= b - \langle \rho(T), \rho_{\rm target} \rangle = b - {\rm Tr}(\rho(T) \rho_{\rm target}) \to \inf,
\label{overlap_to_be_maximized} \\
J_2(c) &= \|\rho(T) - \rho_{\rm target}\|^2 = {\rm Tr}\left( (\rho(T) - \rho_{\rm target})^2 \right) \to \inf, 
\label{steering_to_target_density_matrix} 
\end{align}
where the Hilbert--Schmidt distance is $\| \rho - \sigma\| = \left[{\rm Tr}\left( (\rho - \sigma)^2 \right) \right]^{1/2}$, $\rho(T)$ is the system's final state for certain  control~$c$, and $b$ is the overlap's upper bound obtained as the maximum value of $\langle \rho, \rho_{\rm target}\rangle$, i.e. $b$ is the largest eigenvalue of $\rho_{\rm target}$ (in~\cite{MorzhinPechenQIP2023}, we consider such the problem for the two-qubit case).

We use the representation (parameterization)  
\begin{align}
\rho = \begin{pmatrix}
\rho_{11} & \rho_{12} & \rho_{13} \\ 
\rho_{12}^{\ast} & \rho_{22} & \rho_{23} \\
\rho_{13}^{\ast} & \rho_{23}^{\ast} & \rho_{33}
\end{pmatrix} = 
\begin{pmatrix}
x_1 & x_2 + i x_3 & x_4 + i x_5 \\ 
x_2 - i x_3 & x_6 & x_7 + i x_8 \\
x_4 - i x_5 & x_7 - i x_8 & x_9
\end{pmatrix}, \quad x_j \in \mathbb{R}, \quad j = \overline{1,9}. 
\label{parameterization_of_rho}
\end{align}
After taking $\rho(t)$, $x_1(t), \dots, x_9(t)$ instead of $\rho$, $x_1, \dots, x_9$ and substituting (\ref{parameterization_of_rho}) in (\ref{GKSL_system}), we obtain the bilinear system with real states, 
\begin{align}
\dot x(t) = \Big( A + B^u u(t) + \sum\nolimits_{j=1}^2 B^{n_j} n_j(t) \Big) x(t), \quad x(0) = x_0.
\label{realificated_qutrit_system}
\end{align}
Consider the corresponding 9 differential equations (derived via Wolfram Mathematica), where we set $V_{13} = V_{13}^{\ast}$, $V_{23} = V_{23}^{\ast}$:
\begin{align*}
\dot x_1 &= 2 C_{13} V_{13}^2 x_9 - 2 V_{13} x_5 u 
+ 2 C_{13} V_{13}^2 (x_9 - x_1) n_1, \quad x_1(0) = \rho_{11}(0), \\
\dot x_2 &= - E_2 x_3  - (V_{23} x_5 + V_{13} x_8) u - C_{13} V_{13}^2 x_2 n_1 - C_{23} V_{23}^2 x_2 n_2, 
\quad  x_2(0) = {\rm Re} \, \rho_{12}(0),  \\
\dot x_3 &= E_2 x_2 + (V_{23} x_4 - V_{13} x_7) u 
- C_{13} V_{13}^2 x_3 n_1 - C_{23} V_{23}^2 x_3 n_2, \quad  x_3(0) = {\rm Im} \, \rho_{12}(0),  \\
\dot x_4 &= - E_3 x_5 - (C_{13} V_{13}^2 + C_{23} V_{23}^2) x_4 - V_{23} x_3 u - 2 C_{13} V_{13}^2 x_4 n_1 - C_{23} V_{23}^2 x_4 n_2, \nonumber \\
&\quad x_4(0) = {\rm Re} \, \rho_{13}(0),  \\ 
\dot x_5 &= E_3 x_4 - (C_{13} V_{13}^2 + C_{23} V_{23}^2) x_5 + (V_{13} x_1 - V_{13} x_9 + V_{23} x_2) u - \nonumber \\& \quad 
-2 C_{13} V_{13}^2 x_5 n_1 - C_{23} V_{23}^2 x_5 n_2, \quad x_5(0) = {\rm Im} \, \rho_{13}(0),  \\
\dot x_6 &= 2 C_{23} V_{23}^2 x_9 - 2 V_{23} x_8 u + 2 C_{23} V_{23}^2 (x_9 - x_6) n_2, \quad x_6(0) = \rho_{22}(0),   \\
\dot x_7 &= - (C_{13} V_{13}^2 + C_{23} V_{23}^2) x_7 + (E_2 - E_3) x_8 + V_{13} x_3 u - C_{13} V_{13}^2 x_7 n_1 - \nonumber \\
&\quad - 2 C_{23} V_{23}^2 x_7 n_2, 
\quad x_7(0) = {\rm Re} \, \rho_{23}(0), \\ 
\dot x_8 &= (E_3 - E_2) x_7 - (C_{13} V_{13}^2 + C_{23} V_{23}^2) x_8 + (V_{13} x_2 + V_{23} x_6 -  V_{23} x_9) u - \nonumber \\ &\quad - C_{13} V_{13}^2 x_8 n_1 - 2 C_{23} V_{23}^2 x_8 n_2, \quad x_8(0) = {\rm Im} \, \rho_{23}(0), \\ 
\dot x_9 &= - 2 (C_{13} V_{13}^2 + C_{23} V_{23}^2) x_9 + 2 (V_{13} x_5 + V_{23} x_8) u + 2 C_{13} V_{13}^2 (x_1 - x_9) n_1 + \nonumber \\&\quad + 2 C_{23} V_{23}^2 (x_6 - x_9) n_2, \quad x_9(0) = \rho_{33}(0).
\end{align*}

For (\ref{overlap_to_be_maximized}) and (\ref{steering_to_target_density_matrix}), use of  (\ref{parameterization_of_rho}) gives  the realifications
\begin{align}
J_1(c) &= b - \langle x(T), \beta \circ x_{\rm target}\rangle = b - 
\sum\nolimits_{j=1}^9 \beta_j \, x_{{\rm target},j} \, x_j(T)  \to \inf, 
\label{overlap_to_be_maximized_realificated} \\
J_2(c) &= \sum\nolimits_{j=1}^9 \beta_j \, (x_j(T) - x_{{\rm target},j})^2 \to \inf,
\label{steering_to_target_density_matrix_realificated}
\end{align}
where $\beta=(\beta_j)_{j=1}^9 = (1,2,2,2,2,1,2,2,1)$; ``$\circ$'' denotes the Hadamard product; the vector
\begin{align*}
x_{\rm target} &= 
(\rho_{{\rm target},11}, ~
{\rm Re} \, \rho_{{\rm target},12}, ~
{\rm Im} \, \rho_{{\rm target},12}, ~
{\rm Re} \, \rho_{{\rm target},13}, ~
{\rm Im} \, \rho_{{\rm target},13}, \\
&\qquad 
\rho_{{\rm target},22}, ~
{\rm Re} \, \rho_{{\rm target},23}, ~
{\rm Im} \, \rho_{{\rm target},23}, ~
\rho_{{\rm target},33} ).
\end{align*}

\section{Basic Constructions and Iterative Methods}
\label{Section4}

\subsection{Pontryagin Function, Adjoint System}
\label{Subsection4.1}

For unification, we introduce the objective functional $I(c)$ to be minimized 
and equal either $J_1$ or $J_2$.
For the realificated quantum optimal control problems, we, 
using the general theory of optimal control~\cite{PontryaginBook1962, DemyanovRubinovBook1970}, 
construct the following objects which are necessary for numerical optimization. 
The Pontryagin function is
\begin{align*}
h(x, y, u, n_1, n_2) = \Big\langle y, \Big( A + B^u u + \sum\nolimits_{j=1}^2 B^{n_j} n_j \Big) x \Big\rangle, \quad x,y \in \mathbb{R}^9.
\end{align*} 

Below we consider index $k$ in view of the further consideration of the iterative methods. 

\begin{statement} (adjoint system). For any $J_i$, the general form of the adjoint system at some control process $(x^{(k)}, c^{(k)})$ is the following:
\begin{align}
\dot y^{(k)} &= -\Big( A^{\rm T} + (B^u)^{\rm T} u^{(k)}  + \sum\nolimits_{j=1}^2 (B^{n_j})^{\rm T} n_j^{(k)} \Big) y^{(k)}, 
\label{diff_eq_adjoint_system} \\
y^{(k)}(T) &= -{\rm grad}\,F(x)|_{x = x^{(k)}(T)}.
\label{transversality_condition}
\end{align} 
$F(x)$ is $F_1(x) = b - \sum\nolimits_{j=1}^9 \beta_j x_{{\rm target},j} x_j$ or 
$F_2(x) =  \sum\nolimits_{j=1}^9 \beta_j (x_j - x_{{\rm target},j})^2$. Thus, 
\begin{align*}
y_j^{(k)}(T) = \begin{cases}
\beta_j \, x_{{\rm target},j}, & \text{if} \quad F = F_1, \\
-2 \beta_j \, (x_j^{(k)}(T) - x_{{\rm target},j}), &  \text{if} \quad F = F_2, 
\end{cases} 
\qquad j = 1, \dots, 9. 
\end{align*} 
\end{statement}

Below we give the corresponding 9 differential equations (derived via Wolfram Mathematica) 
taking $V_{13} = V_{13}^{\ast}$, $V_{23} = V_{23}^{\ast}$:
\begin{align*}
\dot y_1^{(k)} &= -V_{13} y_5^{(k)} u^{(k)} - 2 C_{13} V_{13}^2 (y_9^{(k)} - y_1^{(k)}) n_1^{(k)}, \\
\dot y_2^{(k)} &= - E_2 y_3^{(k)} - (V_{13} y_8^{(k)} + V_{23} y_5^{(k)}) u^{(k)} + 
C_{13} V_{13}^2 y_2^{(k)} n_1^{(k)} + C_{23} V_{23}^2 y_2^{(k)} n_2^{(k)}, \\
\dot y_3^{(k)} &= E_2 y_2^{(k)} + (V_{23} y_4^{(k)} - V_{13} y_7^{(k)}) u^{(k)} 
+ C_{13} V_{13}^2 y_3^{(k)} n_1^{(k)} + C_{23} V_{23}^2 y_3^{(k)} n_2^{(k)}, \\
\dot y_4^{(k)} &= (C_{13} V_{13}^2 + C_{23} V_{23}^2) y_4^{(k)} - E_3 y_5^{(k)} - V_{23} y_3^{(k)} u^{(k)} + 
2 C_{13} V_{13}^2 y_4^{(k)} n_1^{(k)} + C_{23} V_{23}^2 y_4^{(k)} n_2^{(k)}, \\
\dot y_5^{(k)} &= (C_{13} V_{13}^2 + C_{23} V_{23}^2) y_5^{(k)} + E_3 y_4^{(k)}  
+ (2 V_{13} (y_1^{(k)} - y_9^{(k)}) + V_{23} y_2^{(k)}) u^{(k)} + \\ 
&\quad + 2 C_{13} V_{13}^2 y_5^{(k)} n_1^{(k)} + C_{23} V_{23}^2 y_5^{(k)} n_2^{(k)}, \\
\dot y_6^{(k)} &= -V_{23} y_8^{(k)} u^{(k)} + 2 C_{23} V_{23}^2 (y_6^{(k)} - y_9^{(k)}) n_2^{(k)}, \\
\dot y_7^{(k)} &= (C_{13} V_{13}^2 + C_{23} V_{23}^2) y_7^{(k)} + (E_2 - E_3) y_8^{(k)} + V_{13} y_3^{(k)} u^{(k)} + \\
&\quad + C_{13} V_{13}^2 y_7^{(k)} n_1^{(k)} + 2 C_{23} V_{23}^2 y_7^{(k)} n_2^{(k)}, \\
\dot y_8^{(k)} &= (E_3 - E_2) y_7^{(k)} + (C_{13} V_{13}^2 + C_{23} V_{23}^2) y_8^{(k)} + 
(V_{13} y_2^{(k)} + 2 V_{23} (y_6^{(k)} - y_9^{(k)})) u^{(k)} + \nonumber \\ 
&\quad + C_{13} V_{13}^2 y_8^{(k)} n_1^{(k)} + 2 C_{23} V_{23}^2 y_8^{(k)} n_2^{(k)}, \\
\dot y_9^{(k)} &= 2 C_{13} V_{13}^2 (y_9^{(k)} - y_1^{(k)}) + 2 C_{23} V_{23}^2 (y_9^{(k)} - y_6^{(k)}) 
+ (V_{13} y_5^{(k)} + V_{23} y_8^{(k)}) u^{(k)} + \nonumber \\
&\quad + 2 C_{13} V_{13}^2 (y_9^{(k)} - y_1^{(k)}) n_1^{(k)} + 2 C_{23} V_{23}^2 (y_9^{(k)} - y_6^{(k)}) n_2^{(k)}.
\end{align*}

\subsection{Gradient of $I = J_i$, $i \in \overline{1,2}$}
\label{Subsection4.2}

\begin{statement} (gradient). For the system (\ref{realificated_qutrit_system}) and any objective functional $I(c) = J_i(c)$ to be minimized, $i \in  \overline{1,2}$, its gradient at a~given admissible control $c^{(k)}$ is as follows:
\begin{align}
{\rm grad} \, I(c^{(k)})(t) &= -h_c(x, y, c)|_{x = x^{(k)}(t), \, y = y^{(k)}(t), \, c = c^{(k)}(t)} = -\mathcal{K}^c(y^{(k)}(t), x^{(k)}(t)) = \nonumber \\
&= -\Big(\langle y^{(k)}(t), B^u x^{(k)}(t)\rangle, ~ \langle y^{(k)}(t), B^{n_j} x^{(k)}(t)\rangle,~ j = 1,2 \Big).
\label{gradint_of_I}
\end{align}
Here the functions $x^{(k)},~y^{(k)}$ are, correspondingly, the solutions of the system (\ref{realificated_qutrit_system}) with this control~$c^{(k)}$ and of the system 
(\ref{diff_eq_adjoint_system}), ({\ref{transversality_condition}}) with the control process $(x^{(k)},c^{(k)})$. In details, the switching functions have the forms
\begin{align*}
\mathcal{K}^u(y,x) &= V_{13} [-2 x_5 y_1 - x_8 y_2 - x_7 y_3 + x_1 y_5 - x_9 y_5 + x_3 y_7 + x_2 y_8 + 
    2 x_5 y_9] + \nonumber \\
&\quad + V_{23} [-x_5 y_2 + x_4 y_3 - x_3 y_4 + x_2 y_5 - 2 x_8 y_6 + x_6 y_8 - x_9 y_8 + 
    2 x_8 y_9 ], \\
\mathcal{K}^{n_1}(y,x) &= -C_{13} V_{13}^2 [-2 x_9 y_1 + x_2 y_2 + x_3 y_3 + 2 x_4 y_4 + 2 x_5 y_5 + x_7 y_7 + \\
&\quad + x_8 y_8 + 2 x_1 (y_1 - y_9) + 2 x_9 y_9], \\ 
\mathcal{K}^{n_2}(y,x) &= -C_{23} V_{23}^2 [x_2 y_2 + x_3 y_3 + x_4 y_4 + x_5 y_5 + 
2 x_6 y_6 - 2 x_9 y_6 + 2 x_7 y_7 + \\ 
&\quad + 2 x_8 y_8 - 2 x_6 y_9 + 2 x_9 y_9].
\end{align*} 
\end{statement}

This statement is formulated for our particular case and is based on the corresponding general gradient formula~\cite{DemyanovRubinovBook1970, PolakBook1971} known in theory of optimal control. 

For a~given admissible  $c^{(k)}$ and arbitrary admissible $c$, we consider the increment $I(c) - I(c^{(k)})$ and its representation $\left\langle {\rm grad} \, I(c^{(k)})(t),~ 
c - c^{(k)} \right \rangle_{L^2[0,T]} + r$ using the gradient~(\ref{gradint_of_I}), where $r$ is the residual. 
The class of piecewise continuous controls is a subclass with respect to the class of square-integrable controls ($L^2([0,T]; Q)$) on the same time interval $[0, T]$, where $Q = Q_{\infty}$ or $Q = Q_{\rm comp}$ with respect to~(\ref{required_constraint_for_n_j}), (\ref{additional_constraints_for_controls}). 
Here we use the scalar product of the space $L^2$, when we consider piecewise continuous controls.

\subsection{Various-Step Gradient Projection Methods}
\label{Subsection4.3}

In the theory of optimal control,  there are the various versions of the one-step GPM~\cite{DemyanovRubinovBook1970}. For some optimal control problems for an open one-qubut system with coherent and incoherent controls, in our articles~\cite{MorzhinPechenIJTP2021, MorzhinPechenLJM2019} the adaptations of the one- and two-parameter versions of GPM-1 are given. For some optimal control problems for an open two-qubit system with coherent and incoherent controls, in our  article~\cite{MorzhinPechenQIP2023} the adaptations of GPM-1, GPM-2 are given. 

Now, for minimizing $I$, we, based on \cite{LevitinPolyak1966, DemyanovRubinovBook1970}, formulate the following GPM-1:
\begin{align}
c^{(k+1)}(t) = {\rm Pr}_Q\left(c^{(k)}(t) - \alpha \, {\rm grad}\,I(c^{(k)})(t) \right), \quad \alpha > 0, \quad k \geq 0.
\label{GPM1}
\end{align}
Here $c^{(0)} = (u^{(0)}, n_1^{(0)}, n_2^{(0)})$ is some admissible initial guess; ${\rm Pr}_Q$ is the orthogonal projection which maps any point outside of $Q$ to a closest point in $Q$, and leaves unchanged points in $Q$; the set $Q = Q_{\infty}$ or $Q = Q_{\rm comp}$  is taken. Let the case that $\alpha > 0$ is fixed for the whole number of iterations. Along with some iteration limit, consider the stopping criterion $I(c^{(k)}) \leq \varepsilon_{\rm stop}$. Of course, one can expect that this inequality can be unreachible (if $T$ is too small, etc.). In details, (\ref{GPM1}) has  the form
\begin{align*}
u^{(k+1)}(t) &= {\rm Pr}_{[-\mu,\mu]}\left( u^{(k)}(t) + \alpha \, \mathcal{K}^u(y^{(k)}(t), x^{(k)}(t)) \right) = \\
&= \begin{cases}
-\mu, & u^{(k)}(t) + \alpha \, \mathcal{K}^u(y^{(k)}(t), x^{(k)}(t)) < -\mu, \\
\mu, & u^{(k)}(t) + \alpha \, \mathcal{K}^u(y^{(k)}(t), x^{(k)}(t)) > \mu, \\
u^{(k)}(t) + \alpha \, \mathcal{K}^u(y^{(k)}(t), x^{(k)}(t)), & |u^{(k)}(t) + \alpha \, \mathcal{K}^u(y^{(k)}(t), x^{(k)}(t))| \leq \mu,
\end{cases} \\
n_j^{(k+1)}(t) &= {\rm Pr}_{[0,n_{\max}]}\left( n_j^{(k)}(t) + \alpha \, \mathcal{K}^{n_j}(y^{(k)}(t), x^{(k)}(t)) \right) = \\
&= \begin{cases}
0, & n_j^{(k)}(t) + \alpha \, \mathcal{K}^{n_j}(y^{(k)}(t), x^{(k)}(t)) < 0, \\
n_{\max}, & n_j^{(k)}(t) + \alpha \, \mathcal{K}^{n_j}(y^{(k)}(t), x^{(k)}(t)) > n_{\max}, \\
n_j^{(k)}(t) + \alpha \, \mathcal{K}^{n_j}(y^{(k)}(t), x^{(k)}(t)), & n_j^{(k)}(t) + \alpha \, \mathcal{K}^{n_j}(y^{(k)}(t), x^{(k)}(t)) \in [0, n_{\max}].
\end{cases}
\end{align*}

Further, based on the works about the heavy-ball method, etc.~\cite{PolyakUSSRComputMathMathPhys1964, PolyakBook1987, AntipinDifferEqu1994, VasilievNedic1994}, we formulate GPM-2 as follows. For a~given initial guess $c^{(0)}$, we construct control $c^{(1)}$ using (\ref{GPM1}). If $k \geq 1$, then GPM-2 has the form
\begin{align}
c^{(k+1)}(t) &= {\rm Pr}_Q\Big(c^{(k)}(t) - \alpha \, {\rm grad} \, I(c^{(k)})(t) + \beta \, (c^{(k)}(t) - c^{(k-1)}(t)) \Big).
\label{GPM2}
\end{align} 
Let the case that $\alpha,~\beta > 0$ are fixed for the whole number of iterations. This method is some infinite-dimensional projection version of the finite-dimensional heavy-ball method. For the last one, the following characteristic is done in the book~\cite[p.~351]{StrangBook2019} by the MIT Math. Prof. G.~Strang: ``So we add momentum with coefficient $\beta$ to the gradient (Polyak's important idea)''. 

Based on \cite{NedichIzvVUZov1993, VasilievNedic1993}, we formulate GPM-3 as follows. For a~given initial guess $c^{(0)}$, we construct control $c^{(1)}$ using~(\ref{GPM1}); after that, we construct control $c^{(2)}$ using~(\ref{GPM2}). If $k \geq 2$, then GPM-3 has the form
\begin{align}
c^{(k+1)}(t) &= {\rm Pr}_Q\Big(c^{(k)}(t) - \alpha \, {\rm grad} \, I(c^{(k)})(t) + 
\beta \, (c^{(k)}(t) - c^{(k-1)}(t)) + \nonumber \\
&\quad + \theta \, (c^{(k-1)}(t) - c^{(k-2)}(t)) \Big).  
\label{GPM3}
\end{align}
Let the case that $\alpha,~\beta,~\theta > 0$ are fixed for the whole number of iterations. 

As Sec.~\ref{Section5} shows below, GPM-2 and GPM-3 can be significantly faster than GPM-1 for the same initial guess and~$\alpha$. 
To the best of our knowledge, use of GPM-3 is new in quantum control.

\subsection{Regularized First-Order Krotov Method}
\label{Subsection4.4}

Based on~\cite{KrotovBook1996, KrotovFeldmanIzvAN1983, MorzhinPechenUMN2019, SrochkoVA_Book2000}, we adapt the first-order RKM for the problem 
$I(c) = J_1(c) \to \inf$:
\begin{align}
c^{(k+1)}(t) = {\rm Pr}_Q\left(c^{(k)}(t) + \alpha \, \mathcal{K}^c(y^{(k)}(t), \underline{x^{(k+1)}(t)}) \right), \quad \alpha > 0, \quad k \geq 0,
\label{RegKrotov1}
\end{align}
where $x^{(k+1)}$ is the solution of the system~(\ref{realificated_qutrit_system}) taken with control $c^{(k+1)}$ to be constructed. Comparing (\ref{GPM1}) and (\ref{RegKrotov1}), we note that GPM-1 uses the gradient (\ref{gradint_of_I}), 
but RKM uses $(-\mathcal{K}^c(y^{(k)}(t), x^{(k+1)}(t)))$ instead of the gradient. In this sense, the method~(\ref{RegKrotov1}) is implicit, in contrast to the explicit method~(\ref{GPM1}). For realizing~(\ref{RegKrotov1}), we find the solution $y^{(k)}$ of  the system (\ref{diff_eq_adjoint_system}), (\ref{transversality_condition}), which takes the process $(x^{(k)}, c^{(k)})$, and solve the special system
\begin{align*}
\dot x^{(k+1)}(t) = \Big( A + B^u u^{\alpha}(x^{(k+1)}(t),t) + \sum\nolimits_{j=1}^2 B^{n_j} n_j^{\alpha}(x^{(k+1)}(t),t) \Big) x^{(k+1))}(t), \quad x(0) = x_0, 
\end{align*}
where we use the projection mappings
$u^{\alpha}(x,t) = {\rm Pr}_{[-\mu,\mu]} \left(u^{(k)}(t) + \alpha \, \mathcal{K}^u(y^{(k)}(t), x) \right)$ and $n_j^{\alpha}(x,t) = {\rm Pr}_{[0, n_{\max}]}\left(n_j^{(k)}(t) + \alpha \, \mathcal{K}^{n_j}(y^{(k)}(t), x) \right)$, $j=1,2$. We use (\ref{RegKrotov1}) with $\alpha>0$ fixed for the whole number of iterations.  

This RKM version is related, first, to the regularized functional $I^{\alpha}(c;c^{(k)}) = I(c) + \frac{1}{2\alpha} \int\nolimits_0^T \| c(t) - c^{(k)}(t) \|^2 dt$ to be minimized, which is considered according to \cite[p.~61]{SrochkoVA_Book2000}, and, second, to the Krotov Lagrangian taken in the form $L^{\alpha}(c, x) = G(x(T)) - \int_0^T R^{\alpha}(t, x(t), c(t))dt$, $G(x(T)) = b - \langle x(T), \beta \circ x_{\rm target} \rangle + 
\langle y(T), x(T) \rangle - \langle y(0), x_0 \rangle$, 
$R^{\alpha}(t, x, c) = \langle y(t), ( A + B^u u + \sum\nolimits_{j=1}^2 B^{n_j} n_j ) x \rangle + 
\langle \dot y(t), x \rangle - \frac{\widehat{\alpha}}{2\alpha}\|c - c^{(k)}(t)\|^2_2$. As~in \cite{SrochkoVA_Book2000}, etc., we note that (\ref{RegKrotov1}) provides $c^{(k+1)}$ such that $I(c^{(k+1)}) \leq I(c^{(k)})$, i.e. the method cannot give $c^{(k+1)}$ worse than $c^{(k)}$ in this sense. For having this property, a sufficient quality of numerical solving of the Cauchy problems is important. 

\section{Numerical Results}
\label{Section5}

As in \cite{PechenTannor2012IsraelJChem}, 
consider $E_2 = 1$, $E_3 = 2.5$, $V_{13} = V_{13}^{\ast} = 1$, 
$V_{23} = V_{23}^{\ast} = 1.7$. For numerical simulations, we use 
the Python programs written by the first author with involving 
such tools as {\tt solve\_ivp} from {\tt SciPy}, etc. 
Controls $u, n_1, n_2$ are interpolated as piecewise constant 
at the uniform time grid introduced by dividing $[0,T]$ into $N$ parts.

In this article, we do not look for such algorithmic parameters of GPM and RKM 
that are the best possible in terms of the methods' complexity 
for a certain problem. The methods' complexity is measured in the number of solved Cauchy problems (our work \cite{MorzhinPechenIrkutsk} also uses this indicator). For each of the considered problems, 
we try to fix such values for the algorithmic parameters for the whole number of
iterations that provide confident decreasing of the objective to be minimized, not necessarily monotonous decrease in $I$ to be minimized, and provide $I(c) \leq \varepsilon_{\rm stop}$. Because GPM-2 has less number of parameters to be adjusted than GPM-3, then it seems to be reasonable to run firstly GPM-2. 

Relying on the various known computational facts 
about the heavy-ball method, e.g.,~\cite{SutskeverMartensDahlHinton2013, TensorFlow_MomentumOptimizer}, 
we, by analogy, take $\beta \in (0, 1)$ and more likely $\beta = 0.8, 0.9$ for GPM-2, GPM-3, but not $\beta = 10$, etc. In {\tt TensorFlow MomentumOptimizer} \cite{TensorFlow_MomentumOptimizer}, such the parameter is set equal to 0.9 by default in the heavy-ball method's implementation. 

\subsection{Steering to a Given $\rho_{\rm target}$ when $\rho_0$ and $\rho_{\rm target}$ have Different Spectra ($J_2 \to \inf$)}
\label{Subsect5.1}

Consider $C_{13} = 0.5$, $C_{23} = 0.3$, $\rho_0 = {\rm diag}(0.8, 0, 0.2)$, 
$\rho_{\rm target} = {\rm diag}(0.5, 0.3, 0.2)$, and $T = 0.5$. 
For the constraints (\ref{additional_constraints_for_controls}), take  $\mu = 50$ and $n_{\max} = 10$. Take $N = 10^3$. The initial guess $c^{(0)} = (1, 0, 0)$ gives $J_2(c^{(0)}) \approx 0.2$. In GPM, consider $\alpha = 1$, $\beta = 0.75$, and $\theta = 0.1$. The condition $J_2 \leq  \varepsilon_{\rm stop} = 10^{-6}$ is satisfied using independently the three versions of GPM, but the numbers of the solved Cauchy problems in these methods are essentially different that is shown in the left subtable in Table~\ref{Table1}. Thus, taking the results of more than one previous iterations can be essentially useful. 

\begin{figure}[ht!]
\centering
\includegraphics[width=1\linewidth]{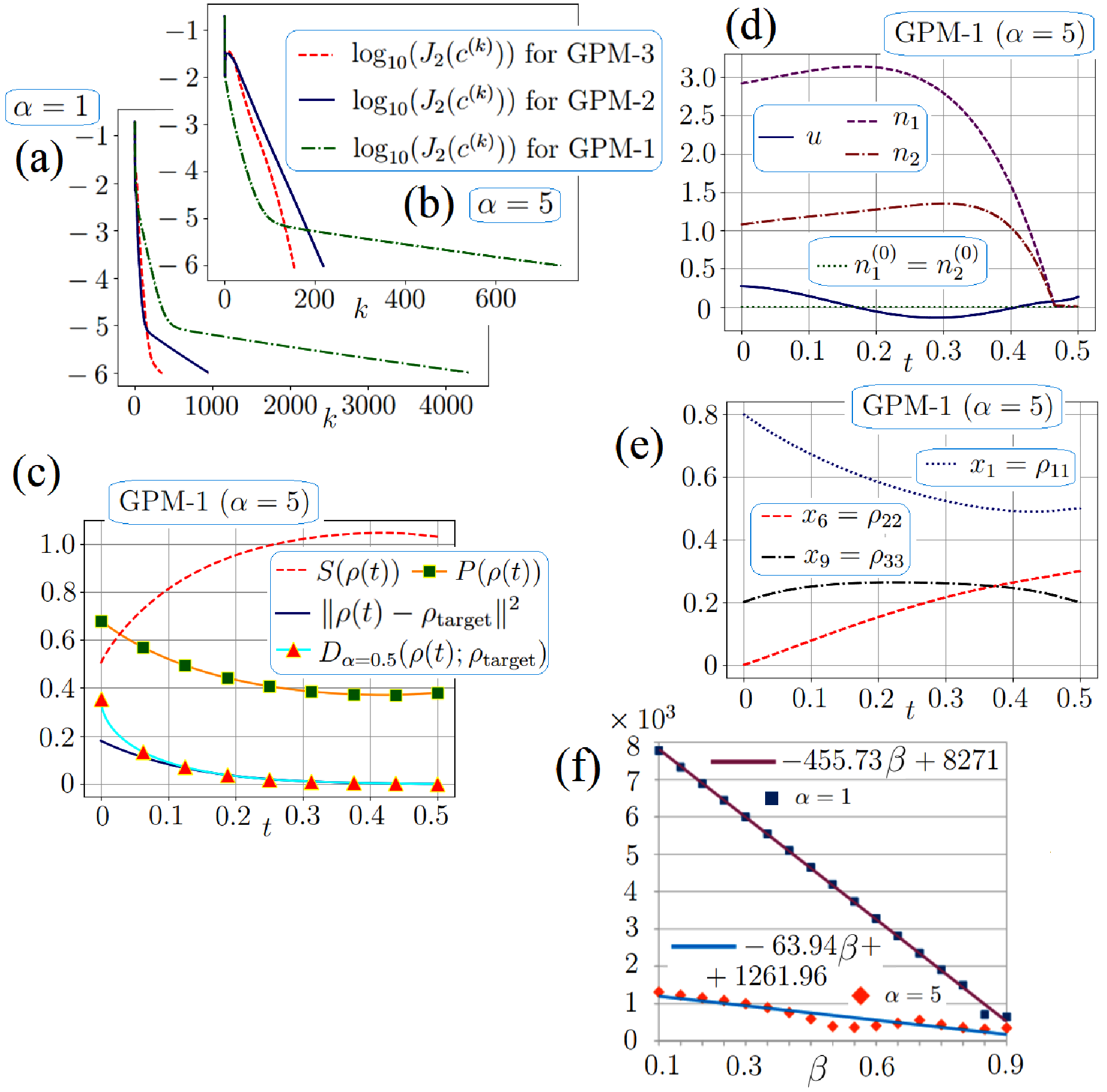}
\caption{For Subsec. 5.1. The results of GPM-1, GPM-2, and GPM-3 for the steering problem: (a,b)~comparing the decreasing rates in the values of~$J_2$, in $\log_{10}$ scale; (c--e)~the optimized controls and describing the corresponding dynamics; (f)~comparing the complexity, as the solved Cauchy problems' numbers, of GPM-2 with $\alpha=1$ and $\alpha=5$ for $\beta = 0.1 + 0.05j$, $j=\overline{0,16}$.}
\label{MorzhinPechenFig1}
\end{figure} 

Further, for the same settings, we change only $\alpha$ by increasing it from 1 to~5. Then the versions of GPM have the complexity
shown in the right subtable in Table~\ref{Table1}. Thus, the increased $\alpha$ decreases the methods' complexities. At the same time, the methods' ranking by their complexity remains the same. 

For the both cases with $\alpha=1$ and $\alpha=5$ and for each method, subplots (a,b) in 
Fig.~\ref{MorzhinPechenFig1} show the corresponding 6~sequences of the values $\log_{10}(J_2(c^{(k)}))$ vs~$k$. Subplots (d,e) show the numerically optimized controls $u, n_1, n_2$ and the corresponding $\rho_{jj}$ ($j=1,2,3$). Subplot (c)  shows $\| \rho(t) - \rho_{\rm target}\|^2$, the von Neumann entropy $S(\rho(t)) = -{\rm Tr}(\rho(t)\log(\rho(t)))$,
purity $P(\rho(t)) = {\rm Tr} (\rho(t))^2$ applied for $\rho(t)$ 
vs~$t$, and also the Petz--R\'{e}nyi $\alpha$-relative entropy 
\cite{Wilde_Book_2017} applied for $\rho(t)$ vs~$t$ and $\rho_{\rm target}$, i.e. $D_{\alpha}(\rho(t); \rho_{\rm target}) = \frac{1}{\alpha - 1} \log {\rm Tr}\left(\rho^{\alpha}(t) \rho^{1-\alpha}_{\rm target} \right) \geq 0$ where $\alpha \in (0,1) \cup (1, \infty)$. 

For these 6~runs, we have the corresponding 6~triples of controls~$u, n_1, n_2$.  Comparing these triples between each other, we note that there are clear differences. Thus, we note the non-uniqueness of solving the control problem for $J_2 \leq 10^{-6}$. 

\begin{table}[h!]
\caption{For Subsec.~5.1. The methods' complexity: how many Cauchy problems were solved for each method. The left subtable is for $\alpha=1$, the right subtable is for $\alpha=5$.}
\vspace{0.15cm}
\label{Table1}
\centering 
 \begin{tabular}{|c|c|}
    \hline
    method & complexity \\ 
    \hline
    GPM-3 & 681 \\ 
    \hline
    GPM-2 & 1907 \\ 
    \hline
    GPM-1 & 8667 \\  
    \hline 
\end{tabular} 
\quad 
\begin{tabular}{|c|c|}
    \hline
    method & complexity \\ 
    \hline
    GPM-3 & 309 \\ 
    \hline
    GPM-2 & 437 \\ 
    \hline
    GPM-1 & 1487 \\  
    \hline 
\end{tabular} 
\end{table} 

In addition, for the steering problem with the same $c^{(0)}$ consider 
GPM-2 with $\alpha=1$ and $\alpha=5$ for $\beta = 0.1 + 0.05j$, $j=\overline{0,16}$. 
In terms of the solved Cauchy problems' numbers, these 34 runs of GPM-2 show various complexity of the method depending on $\alpha,~\beta$. See subplot (f) in 
Fig.~\ref{MorzhinPechenFig1}. For $\alpha=1$ we see that the complexity (square markers) is well and remarkably described by the linear approximation ($-455.73 \beta + 8271$), but for $\alpha=5$ the complexity (rhombic markers) has the noticeable deviations from the linear approximation ($-63.94 \beta + 1261.96$). These graphs show that it is possible to accelerate essentially the method's convergence by increasing $\alpha$ and $\beta$. 

\subsection{Steering to a Given $\rho_{\rm target}$ When $\rho_0$ and $\rho_{\rm target}$ Have the Same Spectrum ($J_2 \to \inf$)}
\label{Subsect5.2}

Consider $C_{13} = C_{23} = 0.5$, $\rho_0 = {\rm diag}(0.1, 0.2, 0.7)$, 
$\rho_{\rm target} = {\rm diag}(0.2, 0.7, 0.1)$, and $T = 0.5$. 
For the constraints (\ref{additional_constraints_for_controls}), 
take  $\mu = 50$ and $n_{\max} = 20$. The value $N = 10^3$. 
The initial guess $c^{(0)} = (0.5, 0, 0)$ gives $J_2(c^{(0)}) \approx 0.02$. In GPM, use $\alpha = 1$, $\beta = 0.75$, and $\theta = 0.1$. The condition $J_2 \leq 10^{-6}$ is satisfied using independently the three versions of GPM, but the numbers of the solved Cauchy problems in these methods are essentially different that is shown in the left subtable in Table~\ref{Table2}. 

Further, for the same settings, we instead of $c^{(0)} = (0.5, 0, 0)$  take $c^{(0)} = (2, 0, 0)$ giving $J_2(c^{(0)}) \approx 0.2$. Here the condition $J_2 \leq 10^{-6}$ is satisfied using independently the three versions of GPM whose complexity is shown in the right subtable in Table~\ref{Table2}.  

\begin{figure}[ht!]
\centering
\includegraphics[width=1\linewidth]{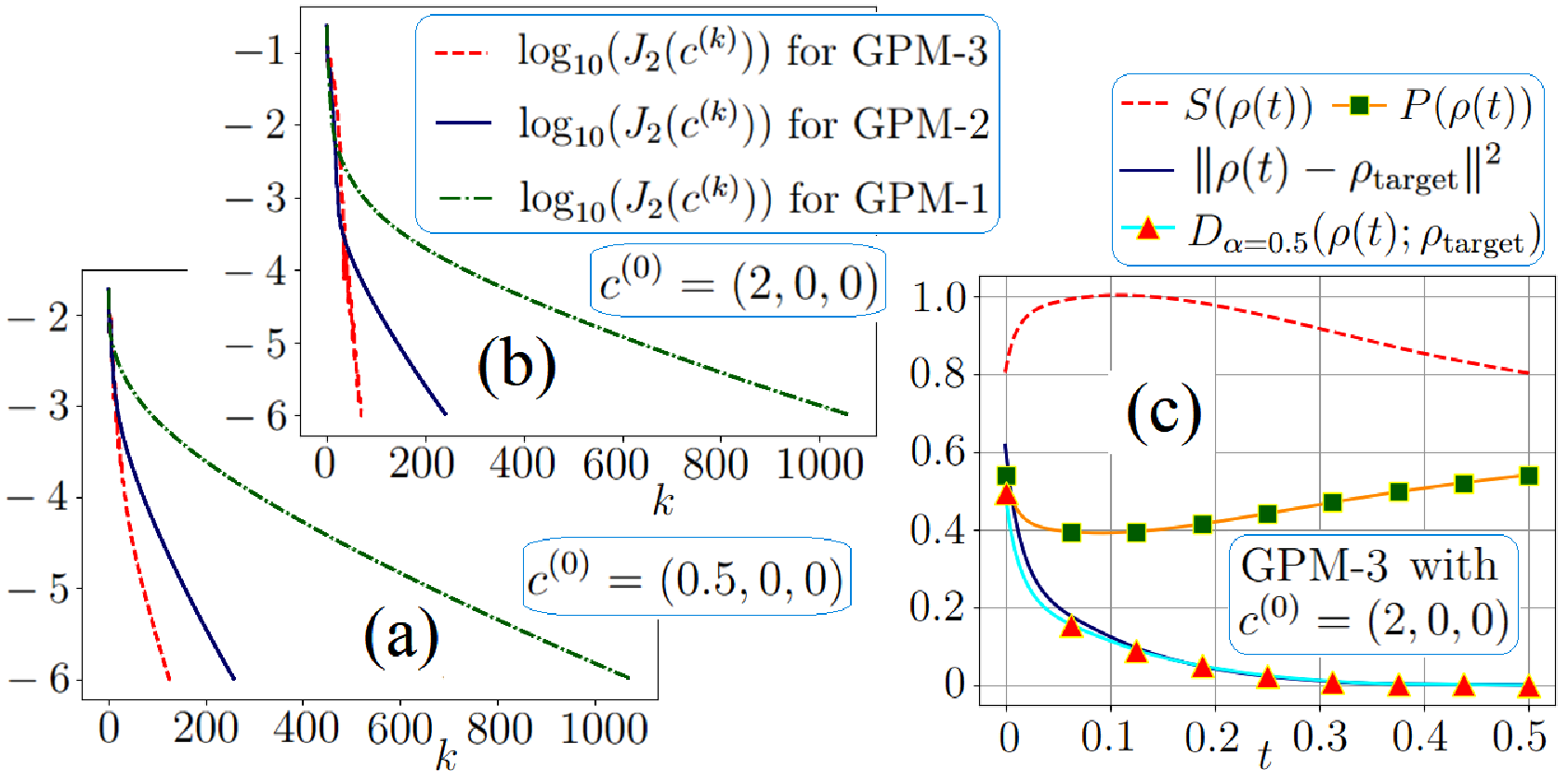}
\caption{For Subsec. 5.2. The results of of GPM-1, GPM-2, and GPM-3 for the steering problem where $\rho_0,~\rho_{\rm target}$ have the same spectrum: (a,b)~comparing the methods' decreasing rates in the values of~$J_2$, in $\log_{10}$ scale; (c)~describing the optimized dynamics.}
\label{MorzhinPechenFig2}
\end{figure} 

For the both cases with $c^{(0)} = (0.5, 0, 0)$ and $c^{(0)} = (2, 0, 0)$ and for each the method, subplots (a,b) in Fig.~\ref{MorzhinPechenFig2} show the corresponding 6~sequences of the values $\log_{10}(J_2(c^{(k)}))$ vs~$k$. For the both initial guesses, the resulting incoherent controls $n_1, n_2$ are non-trivial, despite $n^{(0)} = 0$; the resulting six versions of $u$ are such that $\|u\|_{L^2}$ is close or not close to zero depending on $c^{(0)}$ and which  method is used. Fig.~(c) shows $S(\rho(t))$ vs~$t$ with $S(\rho_0) \approx S(\rho(T))$, also $P(\rho(t))$ vs~$t$ with $P(\rho_0) \approx P(\rho(T))$, decreasing $\| \rho(T) - \rho_{\rm target}\|^2$, and $D_{\alpha=0.5}(\rho(t); \rho_{\rm target})$ which becomes very close to zero at~$t=T$.

\begin{table}[h!]
\caption{For Subsec.~5.2. The methods' complexity: how many Cauchy problems were solved for each method. The left subtable is for $c^{(0)}=(0.5,0.0)$, the right subtable is for $c^{(0)}=(2,0.0)$.}
\vspace{0.15cm}
\label{Table2}
\centering 
 \begin{tabular}{|c|c|}
    \hline
    method & complexity \\ 
    \hline
    GPM-3 & 251 \\ 
    \hline
    GPM-2 & 517 \\ 
    \hline
    GPM-1 & 2151 \\  
    \hline 
\end{tabular} 
\quad 
\begin{tabular}{|c|c|}
    \hline
    method & complexity \\ 
    \hline
    GPM-3 & 141 \\ 
    \hline
    GPM-2 & 485 \\ 
    \hline
    GPM-1 & 2123 \\  
    \hline 
\end{tabular} 
\end{table}

The described above two cases of $c^{(0)}$ and the three versions of GPM provides 6~resulting incoherent controls $n_1$ each of them has such a~part where $n_1(t) > 4$. Take $c = (2, 0, 0)$ and change only the bound $n_{\max}$ by setting $n_{\max} = 4$ instead of $n_{\max} = 20$. GPM-3 with the same $\alpha, \, \beta, \, \theta$ at the cost of solving the 1379 Cauchy problems provides $J_2 \leq \varepsilon_{\rm stop} = 10^{-6}$, i.e. such $c$ is obtained that takes into account the changed constraint and approximately steers to $\rho_{\rm target}$, but the complexity of GPM-3 is essentially higher in comparison to the case with $n_{\max} = 20$ (1379 vs 141 Cauchy problems). Further, we decrease $n_{\max}$ from 4 to~2 and increase $T$ from $T = 0.5$ to $T = 1$. Take the same $N=10^3$. For $c^{(0)} = (2,0,0)$ we use GPM-3 with $\alpha=1$, $\beta = 0.85$, $\theta=0.1$ that gives  such~$c$ that also approximately steers to $\rho_{\rm target}$ in the sense $J_2 \leq \varepsilon_{\rm stop} = 10^{-6}$. The complexity is the 1873 solved Cauchy problems. The $L^2$-norms of the optimized controls $u,~n_1,~n_2$ are not close to zero. These additional experiments remind that the optimization work depends on, in particular, $T$ and the constraints (\ref{additional_constraints_for_controls}).

\subsection{Maximizing the Overlap ($J_1 \to \inf$)}
\label{Subsect5.3}

\begin{figure}[ht!]
\centering
\includegraphics[width=0.95\linewidth]{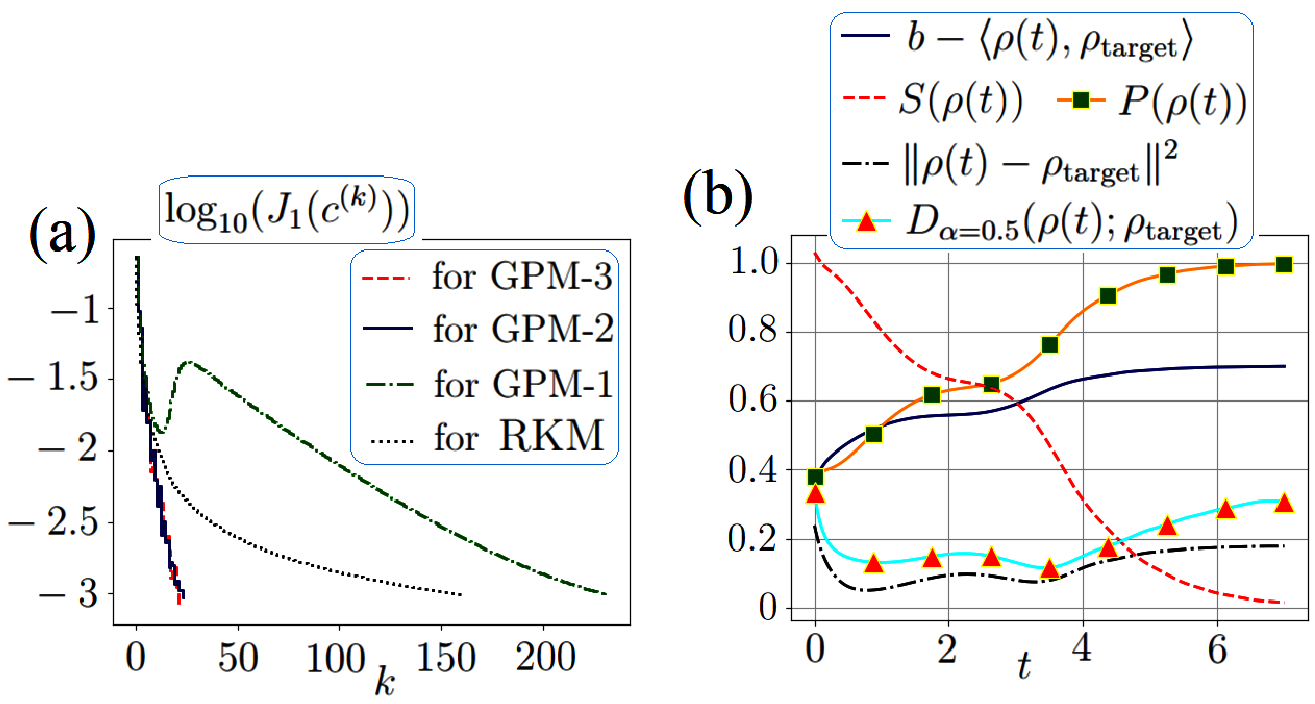}
\caption{For Subsec. 5.3. The numerical results of the various-step GPM and RKM for the problem of minimizing $J_1$: (a)~comparing the methods' decreasing rates in the values of~$J_1$, in $\log_{10}$ scale; (b)~describing the optimized dynamics.}
\label{MorzhinPechenFig3}
\end{figure}

Consider $C_{13} = C_{23} = 0.7$, $\rho_0 = {\rm diag}(0.5, 0.3, 0.2)$, 
$\rho_{\rm target} = {\rm diag}(0.3, 0.7, 0)$, and $T = 7$. 
For the constraints (\ref{additional_constraints_for_controls}), take  $\mu = 50$ and $n_{\max} = 10$. 
The upper bound $b = 0.7$ is equal to the largest eigenvalue 0.7 of $\rho_{\rm target}$ (see \cite[p. 241]{MorzhinPechenQIP2023}). The value $N = 10^3$. The initial guess $c^{(0)} = (0.5, 0, 0)$ gives $J_1(c^{(0)}) \approx 0.2$. In GPM, consider $\alpha = 3$, $\beta = 0.75$, and $\theta = 0.1$. In RKM, $\alpha = 3$ is taken. The condition $J_1 \leq \varepsilon_{\rm stop} = 10^{-3}$ is satisfied using independently the various-step GPM and RKM. The methods' complexity is shown in Table~\ref{Table3}. Thus, GPM-2, GPM-3, and RKM are faster than GPM-1, but GPM-2 is faster than RKM. Thus, the winner is GPM-3 here. For each the method, 
subplot~(a) in Fig.~\ref{MorzhinPechenFig3} shows the corresponding 4~sequences of the values $\log_{10}(J_1(c^{(k)}))$ vs~$k$. 
In subfig.~(a), note the dashed dotted graph showing that GPM-1 with the fixed $\alpha=3$ has --- before the 50th iteration --- firstly significant improvements, then deteriorations, and then again improvements. Subfig.~(b) describes the optimized dynamics (GPM-3). Here $b - \langle \rho(t), \rho_{\rm target} \rangle$ vs~$t$ reaches approximately the upper bound $b = 0.7$ at $t=T$ when $S(\rho(t))$ and $P(\rho(t))$ vs~$t$ reaches, correspondingly, 0 and~1. Because $\rho_{\rm target}$ is a~mixed quantum state, then it is expected to see that the both $\| \rho(T) - \rho_{\rm target}\|^2$ and $D_{\alpha=0.5}(\rho(T); \rho_{\rm target})$ are not close to~0. Each of the four methods give that the resulting $u$ and $n_1$ are non-trivial (although $n^{(0)} = 0$), but with some differences depending on the method. 
 
\begin{table}[h!]
\caption{For Subsec.~5.3. The methods' complexity: how many Cauchy problems were solved for each method.}
\vspace{0.15cm}
\label{Table3}
\centering  
\begin{tabular}{|c|c|c|c|c|}
    \hline
    method & GPM-3 & GPM-2 & GPM-1 & RKM \\ 
    \hline
    complexity & 43 & 47 & 459 & 315 \\ 
    \hline 
\end{tabular}
\end{table}

\section{Conclusions}
\label{Section_Conclusion}
 
The article studied an open qutrit system whose evolution 
is governed by the Gorini--Kossakowski--Sudarshan--Lindblad master equation
with simultaneous coherent and incoherent controls considered as piecewise continuous and bounded, where the initial state $\rho_0$ is given. We considered the problem 
of maximizing the Hilbert--Schmidt overlap between the system's final state $\rho(T)$ and a~given target state $\rho_{\rm target}$ and the problem of minimizing the squared Hilbert--Schmidt distance between these states. For the both problems, we performed their realifications, formulate the corresponding Pontryagin function, adjount system (with the two cases of transversality conditions in view of the two terminal objectives), and gradients of the objectives, adapted the one-, two-, three-step gradient projection methods (GPM-1, GPM-2, and GPM-3). For the problem of maximizing the overlap,  also the regularized first-order Krotov method (RKM) was adapted. Although the performed numerical experiments do not pretend to reflect a~statistical analysis for comparison of the methods, the following observations were made following~Sect.~\ref{Section5}:
\begin{itemize}
    \vspace{-0.3cm}
    \item ranking the methods in terms of their complexity shows that GPM-2, GPM-3, and RKM are faster than GPM-1, that RKM is slower than GPM-2;
    \vspace{-0.3cm}
    \item the computed resulting incoherent controls are non-trivial;
    \vspace{-0.3cm}
    \item depending on how the optimization works and/or 
changing the final time, controls' constraints, we can obtain essentially different profiles of $u,~n_1,~n_2$ providing the same precision~$\varepsilon_{\rm stop}$.
    \vspace{-0.1cm}
\end{itemize}


\begin{thebibliography}{99} 

\bibitem{Holevo_Book_2010_2019}
Холево А.С., {\it Квантовые системы, каналы, информация}, МЦНМО, М., 2010, 328~с., 
\url{https://biblio.mccme.ru/node/2225}. \\
Holevo A.S., {\it Quantum Systems, Channels, Information: A~mathematical Introduction}, 
2nd Rev. and Expanded Ed., De Gruyter, Berlin, Boston, 2019, 15+351~pp., \\
\url{https://doi.org/10.1515/9783110642490}

\bibitem{Holevo_Book_2020_2011}
Холево А.С., {\it Вероятностные и статистические аспекты квантовой теории}, 
МЦНМО, М., 2020, 364~с., \url{https://biblio.mccme.ru/node/30536}.\\
Holevo A., {\it Probabilistic and Statistical Aspects of Quantum Theory}, 
Springer, Basel, 2011, XVI+324~pp., \url{https://doi.org/10.1007/978-88-7642-378-9}

\bibitem{Wilde_Book_2017} 
Wilde M.M., {\it Quantum Information Theory}, 
2nd Ed., Cambridge, Cambridge Univ. Press, 2017, 776~pp., 
\url{https://doi.org/10.1017/9781316809976}

\bibitem{KochEPJQuantumTechnol2022}
Koch C.P., Boscain U., Calarco~T., Dirr~G., Filipp~S., 
Glaser~S., Kosloff~R., Montangero~S., Schulte-Herbr\"{u}ggen~T., 
Sugny~D., Wilhelm~F.K., ``Quantum optimal control in quantum technologies. 
Strategic report on current status, visions and goals for research 
in Europe'', {\it EPJ Quantum Technol.}, {\bf 9} (2022), 19, 60~pp., \\
\url{https://doi.org/10.1140/epjqt/s40507-022-00138-x}

\bibitem{ButkovskiyBook_1984_1990}
Бутковский А.Г., Самойленко Ю.И., {\it Управление квантовомеханическими процессами}, 
Наука, М., 1984, 256~с.; \\
англ. пер.: Butkovskiy A.G., Samoilenko Yu.I.,  
{\it Control of Quantum--Mechanical Processes and Systems}, 
Transl. of the book published in Russian in 1984,   
Kluwer Acad. Publ., Dordrecht, 1990, XIV+232~pp. 

\bibitem{TannorBook2007} 
Tannor D.J., {\it Introduction to Quantum Mechanics: 
A~Time Dependent Perspective}, University Science Books, Sausalito, 2007, 656~pp., \\
\url{https://uscibooks.aip.org/books/introduction-to-quantum-mechanics-a-time-dependent-perspective/}

\bibitem{BrifNewJPhys2010} 
Brif C., Chakrabarti R., Rabitz H.,
``Control of quantum phenomena: Past, present and future'', 
{\it New J.~Phys.}, {\bf 12} (2010), 075008, 68~pp., \\
\url{https://doi.org/10.1088/1367-2630/12/7/075008}

\bibitem{DongPetersen2010}
Dong D.Y., Petersen I.R., 
``Quantum control theory and applications: a survey'', {\it IET Control Theory Appl.}, 
{\bf 4} (2010), 2651--2671, 21~pp., \\
\url{https://doi.org/10.1049/iet-cta.2009.0508}

\bibitem{MooreCS2011} 
Moore K.W., Pechen A., Feng X.-J., Dominy J., 
Beltrani V., Rabitz H., ``Universal characteristics of 
chemical synthesis and property optimization'', 
{\it Chem. Sci.}, {\bf 2} (2011), 417--424, 
\url{https://doi.org/10.1039/C0SC00425A}

\bibitem{Gough2012} 
Gough J., ``Principles and applications of quantum 
control engineering'', {\it Phil. Trans. R. Soc.~A}, {\bf 370} (2012), 5241--5258, 18~pp., \\
\url{https://doi.org/10.1098/rsta.2012.0370}

\bibitem{DongWuYuanLiTarn2015}
Dong W., Wu R., Yuan X., Li C., Tarn T.-J.,  
``The modelling of quantum control systems'', {\it Sci. Bull.}, {\bf 60} (2015), 1493--1508, \\
\url{https://doi.org/10.1007/s11434-015-0863-3}

\bibitem{KochJPhysCondensMatter2016}  
Koch C.P., ``Controlling open quantum systems: Tools, achievements, 
and limitations'', {\it J.~Phys.: Condens. Matter},  
{\bf 28} (2016), 213001, \\ \url{https://doi.org/10.1088/0953-8984/28/21/213001}

\bibitem{DAlessandroBook2021} 
D'Alessandro D., {\it Introduction to Quantum Control and Dynamics}, 
2nd Ed., Chapman and Hall/CRC, Boca Raton, 2021, 416~pp., \\
\url{https://doi.org/10.1201/9781003051268}

\bibitem{BaiChenWuAn2021} 
Bai S.-Y., Chen C., Wu H., An J.-H.,  
``Quantum control in open and periodically driven systems'', 
{\it Adv. Phys.~X}, {\bf 6} (2021), 1870559, \\
\url{https://doi.org/10.1080/23746149.2020.1870559}

\bibitem{CastroGiovanniniEtAlNJP2023}
Castro A., De Giovannini U., Sato S.A, H\"{u}bener H., Rubio A., 
``Floquet engineering with quantum optimal control theory'', 
{\it New J. Phys.}, {\bf 25} (2023), 043023, \\
\url{https://doi.org/10.1088/1367-2630/accb05}

\bibitem{KuprovBook2023}
Kuprov I., {\it Spin: From Basic Symmetries to Quantum Optimal Control},  
Springer, Cham, 2023, XIII+395~pp., 
\url{https://doi.org/10.1007/978-3-031-05607-9}

\bibitem{PontryaginBook1962}   
Понтрягин Л.С., Болтянский В.Г., Гамкрелидзе Р.В., Мищенко~Е.Ф., 
{\it Математическая теория оптимальных процессов}, Физматгиз, М., 1961, 391~с.; \\
англ. пер.: Pontryagin L.S., Boltyanskii V.G., Gamkrelidze~R.V., Mishchenko~E.F., 
{\it The Mathematical Theory of Optimal Processes}, Transl. from Russian,  
Intersci. Publ. JohnWiley \& Sons, Inc., New York, London, 1962, VIII+360~pp.

\bibitem{KrotovBook1996}  
Krotov V.F., {\it Global Methods in Optimal Control Theory},  
Marcel Dekker, New York, 1996.

\bibitem{PalaoKosloffPRL2002}
Palao J.P., Kosloff R., 
``Quantum computing by an optimal control algorithm 
for unitary transformations'', {\it Phys. Rev. Lett.}, 
{\bf 89} (2002), 188301, \\ \url{https://doi.org/10.1103/PhysRevLett.89.188301}

\bibitem{BochkinFeldmanLazarevPechenZenchuk2022}
Bochkin G.A., Fel’dman E.B., 
Lazarev I.D., Pechen~А.N., Zenchuk~A.I., 
``Transfer of zero-order coherence matrix along spin-1/2 chain'',  
{\it Quantum Inf. Process.}, {\bf 21} (2022), 261, 
\url{https://doi.org/10.1007/s11128-022-03613-7}

\bibitem{KazakovKrotov1987} 
Казаков В.А., Кротов В.Ф., ``Оптимальное управление резонансным 
взаимодействием света с веществом'', {\it Автомат. и телемех.}, 
4 (1987), 9--15, 7~pp., \url{https://www.mathnet.ru/rus/at4411}; \\
англ. пер.: Kazakov V.A., Krotov V.F., ``Optimal control of resonant interaction 
between light and matter'', {\it Automat. Remote Control}, {\bf 4} (1987), 430--434, 5~pp. 

\bibitem{PeirceDahlehRabitzPRA1988} 
Peirce A.P., Dahleh M.A., Rabitz H., 
``Optimal control of quantum-mechanical systems: 
Existence, numerical approximation, and applications'', 
{\it Phys. Rev.~A}, {\bf 37} (1988), 4950,
\url{https://doi.org/10.1103/PhysRevA.37.4950}

\bibitem{GoughRatiuSmolyanov2021}  
Gough J.E., Ratiu T.S., Smolyanov O.G.,
``Wigner measures and coherent quantum control'', 
{\it Proc. Steklov Inst. Math.}, {\bf 313} (2021), 52--59, \\
\url{https://doi.org/10.1134/S0081543821020061}

\bibitem{Wu_2007_5681}
Wu R., Pechen A., Brif C., Rabitz H., 
``Controllability of open quantum systems with Kraus-map dynamics'', 
{\it J.~Phys.~A}, {\bf 40}:21 (2007), 5681--5693, \\
\url{https://doi.org/10.1088/1751-8113/40/21/015}

\bibitem{Zhang_Saripalli_Leamer_Glasser_Bondar_2022}
Zhang W., Saripalli R., Leamer J., Glasser R., Bondar~D., 
``All-optical input-agnostic polarization transformer via experimental kraus-map control'', 
{\it Eur. Phys. J. Plus}, {\bf 137} (2022), 930, 
\url{https://doi.org/10.1140/epjp/s13360-022-03104-9}

\bibitem{Goerz_NJP_2014}  
Goerz M.H., Reich D.M., Koch~C.P., 
``Optimal control theory for a unitary operation under dissipative evolution'', 
{\it New J. Phys.}, {\bf 16} (2014), 055012, 28~pp., \\
\url{https://doi.org/10.1088/1367-2630/16/5/055012};  \\
arXiv:1312.0111 [quant-ph], \url{https://doi.org/10.48550/arXiv.1312.0111}

\bibitem{FonsecaFanchiniLimaCastelano2022} 
Fonseca M.E., Fanchini F.F., de Lima E., Castelano L.K., 
``Effectiveness of the Krotov method in controlling open quantum systems'', 
arXiv:2208.03114v3 [quant-ph] (2023), 8~pp.,
\url{https://doi.org/10.48550/arXiv.2208.03114}

\bibitem{PechenPRA062102.2006} 
Pechen A., Rabitz H., 
``Teaching the environment to control quantum systems'', 
{\it Phys. Rev.~A}, {\bf 73} (2006), 062102,
\url{https://doi.org/10.1103/PhysRevA.73.062102}

\bibitem{PechenPRA2011}
Pechen A., ``Engineering arbitrary pure and mixed quantum states'', {\it Phys. Rev.~A},
{\bf 84} (2011), 042106, \url{https://doi.org/10.1103/PhysRevA.84.042106}

\bibitem{DannTobalinaKosloffPRA2020}
Dann R., Tobalina A., Kosloff R., 
``Fast route to equilibration'', {\it Phys. Rev.~A}, {\bf 101} (2020), 052102,
\url{https://doi.org/10.1103/PhysRevA.101.052102} 

\bibitem{MorzhinPechenLJM2019}  
Morzhin O.V., Pechen A.N., 
``Maximization of the overlap between 
density matrices for a two-level open quantum system driven 
by coherent and incoherent controls'', {\it Lobachevskii J.~Math.},   
{\bf 40} (2019), 1532--1548, 17~pp., \\
\url{https://doi.org/10.1134/S1995080219100202}

\bibitem{MorzhinPhysPartNucl2020}  
Моржин О.В., Печень А.Н., ``Максимизация критерия Ульмана–Йожи для открытой 
двухуровневой квантовой системы с когерентным и некогерентным управлениями'', 
ЭЧАЯ, {\bf 51}:4 (2020), 484--493, \\ \url{http://www1.jinr.ru/Pepan/v-51-rus.html}; \\
англ. пер.: Morzhin O.V., Pechen A.N., 
``Maximization of the Uhlmann--Jozsa 
fidelity for an open two-level quantum system with coherent 
and incoherent controls'', {\it Phys. Part. Nucl.}, 
{\bf 51} (2020), 464--469, \url{https://doi.org/10.1134/S1063779620040516}

\bibitem{LokutsievskiyJPA2021} 
Lokutsievskiy L., Pechen A., 
``Reachable sets for two-level open quantum systems driven by coherent and incoherent controls'',  
{\it J.~Phys.~A}, {\bf 54} (2021), 395304,\\ \url{https://doi.org/10.1088/1751-8121/ac19f8}

\bibitem{MorzhinPechenIJTP2021} 
Morzhin O.V., Pechen A.N., 
``Minimal time generation of density matrices 
for a two-level quantum system driven by coherent and incoherent controls'',  
{\it Internat. J. Theoret. Phys.}, {\bf 60} (2021), 576--584, 
9~pp.,\\
\url{https://doi.org/10.1007/s10773-019-04149-w}

\bibitem{PetruhanovPhotonics2023} 
Petruhanov V.N., Pechen A.N., 
``Quantum gate generation in two-level open quantum systems by coherent and incoherent photons 
found with gradient search'', {\it Photonics}, {\bf 10} (2023), 220, 15~pp.,
\url{https://doi.org/10.3390/photonics10020220}

\bibitem{MorzhinPechenQIP2023}  
Morzhin O.V., Pechen A.N., 
``Optimal state manipulation for a~two-qubit system driven by coherent and incoherent controls'', 
{\it Quantum Inf. Process.}, {\bf 22} (2023), 241, 26~pp.,
\url{https://doi.org/10.1007/s11128-023-03946-x}

\bibitem{KozyrevPechenPRA2022}
Kozyrev S.V., Pechen A.N., ``Quantum feedback control in quantum photosynthesis'', 
{\it Phys. Rev.~A}, 106 (2022), 32218, \url{https://doi.org/10.1103/PhysRevA.106.032218}

\bibitem{LikhanskiiNapartovich1982}
Лиханский В.В., Напартович А.П., 
``Неадиабатические переходы в трехуровневой системе в поле лазерного излучения 
с плавно меняющейся частотой'', {\it Квантовая электроника}, {\bf 9}:8 (1982), 1591--1599,\\
\url{https://www.mathnet.ru/rus/qe5774}; \\
англ. пер.: Likhanskii~V.V., Napartovich~A.P.,  
``Nonadiabatic transitions in a three--level system 
subjected to a~laser radiation field with a~smoothly varying frequency'', 
{\it Sov.~J. Quantum Electron.}, {\bf 12} (1982), 1020--1024,\\
\url{https://doi.org/10.1070/QE1982v012n08ABEH005774}

\bibitem{LiPengPRA1985}
Li X.-S., Peng Y.-N., 
``Quantum properties of a three-level atom 
interacting with two radiation fields'', {\em Phys. Rev.~A}, {\bf 32} (1985), 1501,\\
\url{https://doi.org/10.1103/PhysRevA.32.1501}

\bibitem{WangMullerEtAl2005}
Wang Q.Q., Muller A., Cheng M.T., Zhou H.J., Bianucci P., Shih C.K., 
``Coherent control of a~$V$-type three--level system in a~single quantum dot'',  
{\it Phys. Rev. Lett.}, {\bf 95} (2005), 187404,
\url{https://doi.org/10.1103/PhysRevLett.95.187404}

\bibitem{AccardiKozyrevPechen2006}
Accardi L., Kozyrev S.V., Pechen A.N., 
``Coherent quantum control of $\Lambda$-atoms 
through the stochastic limit'', In {\it Quantum Information and Computing},  
Edited by L.~Accardi, M.~Ohya, N.~Watanabe,  
World Scientific, Hackensack, NJ, 2006, 1--17,\\ \url{https://doi.org/10.1142/5991}\\
(Book Series: QP-PQ: Quantum Probability and White Noise Analysis; Vol.~19). 
arXiv version (2004),\\ \url{https://doi.org/10.48550/arXiv.quant-ph/0403100}

\bibitem{SugnyKontz2008}   
Sugny D., Kontz C., ``Optimal control of a three-level quantum system 
by laser fields plus von Neumann measurements'',  
{\it Phys. Rev.~A}, {\bf 77} (2008), 063420,\\
\url{https://doi.org/10.1103/PhysRevA.77.063420}

\bibitem{JieYaoMinHaiRui2009}
Jie Z., Yao-Min D., Hai-Rui W.,  
``Realization of two-qutrit quantum gates with control pulses'',  
{\it Commun. Theor. Phys.}, {\bf 51} (2009), 653--658,\\
\url{https://doi.org/10.1088/0253-6102/51/4/15}

\bibitem{ZobovShauro2011}  
Зобов В.Е., Шауро В.П., ``Об оптимальном по времени управлении
методом ЯМР состояниями кутритов, представленных
квадрупольными ядрами со спином $I=1$'', {\it ЖЭТФ}, {\bf 140}:2 (2011), 211--223, 13~pp., \\
\url{http://www.jetp.ras.ru/cgi-bin/r/index/r/140/2/p211?a=list}; \\
англ. пер.: Zobov V.E., Shauro V.P., 
``On time-optimal NMR control of states of qutrits represented by quadrupole 
nuclei with the spin $I = 1$'',  {\it J.~Exp. Theor. Phys.}, {\bf 113} (2011), 181--191,
\url{https://doi.org/10.1134/S1063776111060094}

\bibitem{PechenTannorPRL2011}  
Pechen A.N., Tannor D.J., 
``Are there traps in quantum control landscapes?'', 
{\it Phys. Rev. Lett.}, {\bf 106} (2011), 120402,\\
\url{https://doi.org/10.1103/PhysRevLett.106.120402}

\bibitem{PechenTannor2012PRL} 
Pechen A.N., Tannor D.J., 
``Pechen and Tannor Reply'', {\it Phys. Rev. Lett.}, 
{\bf 108} (2012), 198902, \url{https://doi.org/10.1103/PhysRevLett.108.198902}

\bibitem{PechenTannor2012IsraelJChem}  
Pechen A.N., Tannor D.J., 
``Quantum control landscape for a~Lambda-atom in the vicinity of second-order traps'',  
{\it Israel J. Chem.}, {\bf 52} (2012), 467--472,\\ \url{https://doi.org/10.1002/ijch.201100165};  \\
arXiv: 1508.04169, \url{https://doi.org/10.48550/arXiv.1508.04169}
 
\bibitem{Mortensen2018}   
Mortensen H.L., S{\o}rensen J.J.W.H., M{\o}lmer K., 
Sherson J.F., ``Fast state transfer in a $\Lambda$-system: 
a~shortcut-to-adiabaticity approach to robust 
and resource optimized control'', {\it New J. Phys.}, {\bf 20} (2018), 025009,\\
\url{https://doi.org/10.1088/1367-2630/aaac8a}

\bibitem{DAlessandroShellerZhu2020}  
D'Alessandro D., Sheller B.A., Zhu Z.,
``Time-optimal control of quantum lambda systems 
in the KP configuration'', {\it J.~Math. Phys.}, {\bf 61} (2020), 052107,\\
\url{https://doi.org/10.1063/5.0008034}

\bibitem{PetiziolArimondoEtAl2020}
Petiziol F., Arimondo E., Giannelli L., Mintert F., Wimberger S.,
``Optimized three-level quantum transfers based 
on frequency-modulated optical excitations'',  
{\it Sci. Rep.}, {\bf 10} (2020), 2185, 
\url{https://doi.org/10.1038/s41598-020-59046-8}

\bibitem{DehaghaniLoboPereiraAguiar2022} 
Dehaghani N.B., Lobo Pereira F., Aguiar A.P., 
``A~quantum optimal control problem with state constrained 
preserving coherence'', {\it Proc. 2022 IEEE 61st Conf. CDC} (2022),
\url{https://doi.org/10.1109/CDC51059.2022.9993086}

\bibitem{XuSongWangYe2022}
Xu H., Song X.-K., Wang D., Ye~L., 
``Robust coherent control in three-level quantum systems 
using composite pulses'', {\it Opt. Express},  
{\bf 30} (2022), 3125--3137,\\
\url{https://doi.org/10.1364/OE.449426}

\bibitem{KorashyAxioms2023}
Korashy S., Abdel-Aty M., ``Quantum control of a~nonlinear 
time-dependent interaction of a~damped three-level atom'',  
{\it Axioms}, {\bf 12} (2023), 552, \\
\url{https://doi.org/10.3390/axioms12060552}

\bibitem{LenziGabrick_et_al_QuantumReports2023}
Lenzi E.K., Gabrick E.C., Sayari E., de Castro~A.S.M.,   
Trobia~J., Batista~A.M., ``Anomalous relaxation and three-level system: 
A~fractional Schr\"{o}dinger equation approach'', {\it Quantum Rep.},   
{\bf 5} (2023), 442--458,\\ \url{https://doi.org/10.3390/quantum5020029}

\bibitem{ElovenkovaPechenQuantumReports2023}
Elovenkova M., Pechen A., ``Control landscape of measurement-assisted 
transition probability for a three-level quantum system with dynamical symmetry'',  
{\it Quantum Rep.}, {\bf 5} (2023), 526--545, 20~pp., 
\url{https://doi.org/10.3390/quantum5030035}

\bibitem{KuznetsovPechenLJM2023} 
Kuznetsov S.A., Pechen A.N., 
``On controllability of $\Lambda$- and $V$-atoms and other three-Level systems 
with two allowed transitions'', {\it Lobachevskii J. Math.}, 
{\bf 44}:6 (2023), 2099--2106 (in press). DOI: 10.1134/S199508022306029X

\bibitem{BoscainPRXQuantum2021}  
Boscain U., Sigalotti M., Sugny D.,   
``Introduction to the Pontryagin maximum principle 
for quantum optimal control'', {\it PRX Quantum}, {\bf 2} (2021), 030203,\\
\url{https://doi.org/10.1103/PRXQuantum.2.030203}

\bibitem{Tannor1992}  
Tannor D.J., Kazakov V., Orlov V., 
``Control of photochemical branching: Novel procedures for finding optimal pulses and global 
upper bounds'', In {\it Time-Dependent Quantum Molecular Dynamics},  
Springer, Boston, 1992, 347--360,\\
\url{https://doi.org/10.1007/978-1-4899-2326-4_24}

\bibitem{Schirmer_deFouquieres_2011}
Schirmer S.G., de Fouquieres P., 
``Efficient algorithms for optimal control of quantum dynamics: the Krotov method unencumbered'', 
{\it New J. Phys.}, {\bf 13} (2011), 073029,\\
\url{https://doi.org/10.1088/1367-2630/13/7/073029}

\bibitem{MorzhinPechenUMN2019} 
Моржин О.В., Печень А.Н., 
``Метод Кротова в задачах оптимального управления
замкнутыми квантовыми системами'', {\it УМН}, {\bf 74}:5(449) (2019), 83--144,\\
\url{https://www.mathnet.ru/rus/rm9835}; \\ 
англ. пер.: Morzhin O.V., Pechen A.N., 
``Krotov method for optimal control of closed quantum systems'', 
{\it Russian Math. Surveys}, {\bf 74} (2019), 851--908,\\
\url{https://doi.org/10.1070/RM9835}

\bibitem{ArakiNoriGneitingPRA2023}  
Araki T., Nori F., Gneiting C.,   
``Robust quantum control with disorder-dressed evolution'',  
{\it Phys. Rev.~A}, {\bf 107} (2023), 032609,\\
\url{https://doi.org/10.1103/PhysRevA.107.032609}

\bibitem{MorzhinPechenIrkutsk} 
Моржин О.В., Печень А.Н., 
``Оптимизация когерентного и некогерентного управлений для открытых 
двухкубитных систем'', {\it Изв. Иркутск. гос. ун-та. Сер. ``Математика''}, 
{\bf 45} (2023) (в~печати). (Morzhin O.V., Pechen A.N., 
``Optimization of Coherent and Incoherent Controls for Open Two-Qubit Systems'',
{\it Bull. Irkutsk State Univ. Ser. Math.}, in press).

\bibitem{KrotovFeldmanIzvAN1983}  
Кротов В.Ф., Фельдман И.Н., ``Итерационный метод решения задач оптимального управления'', 
{\it Изв. АН СССР. Техн. киберн.}, 2 (1983), 160--168, 9~с.; \\
англ. пер.: Krotov V.F., Feldman I.N., ``An iterative method for solving problems 
of optimal control'', {\it Engrg. Cybern.}, {\bf 21} (1983), 123--130. 

\bibitem{SrochkoVA_Book2000}
Срочко В.А., {\it Итерационные методы решения задач оптимального управления},
Физматлит, Москва, 2000, 160~с.,\\ \url{https://search.rsl.ru/ru/record/01000686861} \\
(Srochko V.A., {\it Iterative Methods for Solving Optimal Control Problems},
Fizmatlit, Moscow, 2000, in Russian).

\bibitem{LyakhovPechen2022} 
Lyakhov K.A., Pechen A.N., 
``Laser and diffusion driven optimal discrimination of similar 
quantum systems in resonator'', {\it Lobachevskii J. Math.}, {\bf 43}:7 (2022), 1693--1703,
\url{https://doi.org/10.1134/S1995080222100249}

\bibitem{KhanejaJMagnReson2005}  
Khaneja N., Reiss T., Kehlet C., Schulte-Herbr\"{u}ggen T., Glaser S.J.,  
``Optimal control of coupled spin dynamics: 
design of NMR pulse sequences by gradient ascent algorithms'',  
{\it J.~Magn. Reson.}, {\bf 172} (2005), 296--305,\\
\url{https://doi.org/10.1016/j.jmr.2004.11.004}

\bibitem{SchulteHerbruggenSporlKhanejaGlaser2011}  
Schulte-Herbr\"{u}ggen T., Sp\"{o}rl~A.,  
Khaneja~N., Glaser~S.J., ``Optimal control for generating quantum gates 
in open dissipative systems'', {\it J.~Phys.~B}, {\bf 44} (2011), 154013,
\url{https://doi.org/10.1088/0953-4075/44/15/154013}

\bibitem{VolkovJPA2021}  
Volkov B.O., Morzhin O.V., Pechen A.N., 
``Quantum control landscape for ultrafast generation 
of single-qubit phase shift quantum gates'',  
{\it J.~Phys.~A}, {\bf 54} (2021), 215303,
\url{https://doi.org/10.1088/1751-8121/abf45d}

\bibitem{PetruhanovPechenIntJModPhysB2022}  
Petruhanov V.N., Pechen A.N., ``Optimal control for 
state preparation in two-qubit open quantum systems driven 
by coherent and incoherent controls via GRAPE approach'',  
{\it Int.~J. Mod. Phys.~B}, {\bf 37} (2022), 2243017,\\
\url{https://doi.org/10.1142/S0217751X22430175}

\bibitem{GoodwinVinding2023}
Goodwin D.L., Vinding M.S., 
``Accelerated Newton-Raphson GRAPE methods for optimal control'', 
{\it Phys. Rev. Research}, {\bf 5} (2023), L012042,\\
\url{https://doi.org/10.1103/PhysRevResearch.5.L012042}

\bibitem{MorzhinIzvRAN2023}  
Моржин О.В., Печень А.Н., 
``Об оптимизации когерентного и некогерентного управлений для двухуровневых квантовых систем'', 
{\it Изв. РАН. Сер. матем.}, {\bf 87}:5 (2023),
\url{https://doi.org/10.4213/im9372} (в печати); \\
arXiv:2205.02521 [quant-ph], \url{https://doi.org/10.48550/arXiv.2205.02521}

\bibitem{CanevaPRA2011} 
Caneva T., Calarco T., Montangero S., 
``Chopped random-basis quantum optimization'', {\it Phys. Rev.~A}, {\bf 84} (2011), 022326,\\
\url{https://doi.org/10.1103/PhysRevA.84.022326}

\bibitem{MullerSaidJelezkoCalarcoMontangero2022} 
M\"{u}ller M.M., Said R.S., Jelezko F., Calarco T., Montangero S.,  
``One decade of quantum optimal control in the chopped random basis'',  
{\it Rep. Prog. Phys.}, {\bf 85} (2022), 076001,
\url{https://doi.org/10.1088/1361-6633/ac723c}

\bibitem{PechenBorisenokFradkov2022}
Pechen A.N., Borisenok S., Fradkov A.L., 
``Energy control in a quantum oscillator using coherent 
control and engineered environment'', {\it Chaos Solitons Fractals}, 
{\bf 164} (2022), 112687, 8~pp.,
\url{https://doi.org/10.1016/j.chaos.2022.112687}

\bibitem{AndrievskyFradkovAiT2021}
Андриевский Б.Р., Фрадков А.Л., 
``Метод скоростного градиента и его приложения'', 
{\it Автомат. и телемех.}, 9 (2021), 3--72,\\ \url{https://www.mathnet.ru/rus/at15554}; \\
англ. пер.: Andrievsky B.R., Fradkov A.L., 
``Speed gradient method and its applications'', 
{\it Autom. Remote Control}, {\bf 82}:9 (2021), 1463--1518,\\
\url{https://doi.org/10.1134/S0005117921090010}

\bibitem{ShaoCombesHauserNicotraPRA2022}
Shao J., Combes J., Hauser J., Nicotra M.M., 
``Projection-operator-based Newton method for the trajectory 
optimization of closed quantum systems'', {\it Phys. Rev.~A}, {\bf 105} (2022), 032605,
\url{https://doi.org/10.1103/PhysRevA.105.032605}

\bibitem{ShaoNarisHauserNicotra2023}
Shao J., Naris M., Hauser J., Nicotra M.M., 
``How to solve quantum optimal control problems using projection operator-based Newton steps'',  
arXiv:2305.17630 [quant-ph] (2023), \url{https://doi.org/10.48550/arXiv.2305.17630}

\bibitem{LevitinPolyak1966}
Левитин Е.С., Поляк Б.Т., 
``Методы минимизации при наличии ограничений'', 
{\it Ж.~вычисл. матем. и~матем. физ.}, {\bf 6}:5 (1966), 787--823, 37~с.,\\
\url{http://mi.mathnet.ru/zvmmf7415}; \\
англ. пер.: Levitin E.S., Polyak B.T., 
``Constrained minimization methods'', 
{\it USSR Comput. Math. \& Math. Phys.}, {\bf 6} (1966), 1--50,\\
\url{https://doi.org/10.1016/0041-5553(66)90114-5}

\bibitem{DemyanovRubinovBook1970} 
Демьянов В.Ф., Рубинов А.М., ``Приближенные методы решения экстремальных задач'', 
Изд-во Ленинград. ун-та, Л., 1968, 180~с.; \\
англ. пер.: Demyanov V.F., Rubinov A.M., {\it Approximate Methods in Optimization Problems},  
American Elsevier Pub. Co., New York, 1970, IX+256~pp.

\bibitem{PolakBook1971}
Полак Э., {\it Численные методы оптимизации: единый подход}, 
Изд-во ``Мир'', М., 1974, 374~с.; \\
пер. с англ.: Polak E., ``Computational Methods in Optimization: 
A~Unified Approach'', Academic Press, New York, London, 1971.

\bibitem{PolyakUSSRComputMathMathPhys1964} 
Поляк Б.Т., ``О некоторых способах ускорения сходимости итерационных 
методов'', {\it Ж.~вычисл. матем. и матем.~физ.}, 
{\bf 4}:5 (1964), 791--803, 13~с.,\\ \url{http://mi.mathnet.ru/zvmmf7713}; \\
англ. пер.: Polyak B.T., 
``Some methods of speeding up the convergence of 
iteration methods'', {\it USSR Comput. Math. Math. Phys.}, 
{\bf 4} (1964), 1--17,\\ \url{https://doi.org/10.1016/0041-5553(64)90137-5}

\bibitem{PolyakBook1987} 
Поляк Б.Т., {\it Введение в оптимизацию}, Наука, М., 1983, 384~с.; \\
англ. пер.: Polyak B.T., ``Introduction to Optimization'',  
Optimization Software Inc., Publ. Division, New York, 1987.

\bibitem{AntipinDifferEqu1994} 
Антипин А.С., ``Минимизация выпуклых функций на выпуклых множествах
с помощью дифференциальных уравнений'', 
{\it Дифференц. уравнения}, {\bf 30}:9 (1994),
1475--1486, \url{https://www.mathnet.ru/rus/de9433}; \\
англ. пер.: Antipin A.S., 
``Minimization of convex functions on convex sets 
by means of differential equations'', {\it Differ. Equat.},  
{\bf 30} (1994), 1365--1375. 

\bibitem{VasilievNedic1994}
Васильев Ф.П., Недич А., 
``Регуляризованный непрерывный метод проекции градиента второго порядка'', 
{\it Вестн. Моск. ун-та., Сер.~15}, 2 (1994), 3--11; \\
англ. пер.: Vasilev F.P., Nedi\'c A., ``A regularized continuous 
gradient projection method of the second order'',  
{\it Moscow Univ. Comput. Math. Cybernet.},  
{\bf 2} (1994), 1--9. 

\bibitem{NedichIzvVUZov1993}
Недич А., 
``Трехшаговый метод проекции градиента для задач минимизации'', 
{\it Изв. вузов. Матем.}, 10 (1993), 32--37, 6~с.,\\
\url{https://www.mathnet.ru/rus/ivm4551}; \\
англ. пер.: Nedich A., ``The three-step gradient projection method 
for minimization problems'', {\it Russian Math. (Izv. VUZ)}, 
{\bf 37}:10 (1993), 30--36. 

\bibitem{VasilievNedic1993}
Васильев Ф.П., Недич А., 
``О трехшаговом регуляризованном методе проекции градиента 
для решения задач минимизации с неточными исходными данными'', 
{\it Изв. вузов. Матем.}, 12 (1993), 35--43, 9~с.,\\
\url{https://www.mathnet.ru/rus/ivm4577}; \\
англ. пер.: Vasilev F.P., Nedi\'c A., ``A three-step regularized gradient 
projection method for solving minimization problems 
with inexact initial data'', {\it Russian Math. (Izv. VUZ)},   
{\bf 37}:12 (1993), 34--43.   

\bibitem{Kaewyong_Sitthithakerngkiet_Axioms_2021}
Kaewyong N., Sitthithakerngkiet K., 
``A~self-adaptive algorithm for the common solution of the split minimization problem 
and the fixed point problem'', {\it Axioms}, {\bf 10} (2021), 109,
\url{https://doi.org/10.3390/axioms10020109}

\bibitem{KozlovSmolyanov2021}
Козлов В.В., Смолянов О.Г., 
``Математические структуры, связанные с описанием квантовых состояний'', 
{\it Докл. РАН. Мат. информ. проц. упр.}, {\bf 501}:1 (2021), 57--61, 5~с.,
\url{https://www.mathnet.ru/rus/danma20}; \\
англ. пер.: Kozlov V.V., Smolyanov O.G., ``Mathematical structures 
related to the description of quantum states'',  
{\it Dokl. Math.}, {\bf 104} (2021), 365--368. 

\bibitem{GoughOrlovSakbaevSmolyanov2021} 
Гоф Дж., Орлов Ю.Н., Сакбаев В.Ж., Смолянов О.Г., 
``Рандомизированное квантование гамильтоновых систем'', 
{\it Докл. РАН. Матем., информ., проц. упр.}, 498 (2021), 31--36, 6~с.,
\url{https://www.mathnet.ru/rus/danma16}; \\
англ. пер.: Gough J.E., Orlov Yu.N., Sakbaev V.Zh., Smolyanov O.G., 
``Random quantization of Hamiltonian systems'',  
{\it Dokl. Math.}, {\bf 103} (2021), 122--126.  

\bibitem{VacchiniPRE2001} 
Vacchini B., ``Test particle in a quantum gas'', {\it Phys. Rev.~E}, {\bf 63}:6 (2001), 066115,\\
\url{https://doi.org/10.1103/PhysRevE.63.066115}

\bibitem{VacchiniPR2009} 
Vacchini B., Hornberger K., 
``Quantum linear Boltzmann equation'', {\it Phys. Rep.}, {\bf 478}:4--6 (2009), 71--120,
\url{https://doi.org/10.1016/j.physrep.2009.06.001}

\bibitem{StrangBook2019}
Strang G., {\it Linear Algebra and Learning from Data},  
Wellesley--Cambridge Press, 2019, 432~pp. 

\bibitem{SutskeverMartensDahlHinton2013}
Sutskever I., Martens J., Dahl G., Hinton G.,  
``On the importance of initialization and momentum in deep learning'',  
{\it PMLR}, {\bf 28} (2013), 1139--1147, \\
\url{https://proceedings.mlr.press/v28/sutskever13.html}

\bibitem{TensorFlow_MomentumOptimizer}
TensorFlow, machine learning platform: MomentumOptimizer. 
Available online:\\ \url{https://www.tensorflow.org/api_docs/python/tf/compat/v1/train/MomentumOptimizer} \\
(accessed on 5 August, 2023). 

\end{thebibliography}
\end{document}